\newcommand{\be}{\begin{equation}}
\newcommand{\ee}{\end{equation}}
\newcommand{\bea}{\begin{eqnarray}}
\newcommand{\eea}{\end{eqnarray}}
\newcommand{\ba}[1]{\begin{array}{#1}}
\newcommand{\ea}{\end{array}}

\documentclass[twocolumn, prl, amssymb,aps,showpacs,twocolumn,amsmath,showkeys,floatfix]{revtex4}

\usepackage{graphicx}
\usepackage{dcolumn}
\usepackage{bm}
\usepackage{longtable}
\usepackage{epsfig}
\usepackage{times}
\usepackage{commath}
\usepackage{url}
\usepackage{color}

\DeclareMathOperator{\tr}{tr}
\begin{document}
%%%%%%%%%%%%%%%%%%%%%%%%%%%%%%%%%%%%%%%%%%%%%%%%%%%%%%%%%%%%%%%%%%%%%%%%%%%%%%%%
%%%%%%%%%%%%%%%%%%%%%%%%%%%%%%%%%%%%%%%%%%%%%%%%%%%%%%%%%%%%%%%%%%%%%%%%%%%%%%%%

\title{Photonic Controlled-Phase Gates Through Rydberg Blockade in Optical Cavities}

\author{Sumanta \surname{Das}$^{1}$}
\email{sumanta@nbi.ku.dk}

\author{Andrey \surname{Grankin}$^{2}$}
\email{andrey.grankin@u-psud.fr}

\author{Ivan \surname{Iakoupov}$^{1}$}

\author{Etienne \surname{Brion}$^{3}$}

\author{Johannes \surname{Borregaard}$^{1,4}$}

\author{Rajiv \surname{Boddeda}$^{2}$}

\author{Imam  \surname{Usmani}$^{2}$}

\author{Alexei \surname{Ourjoumtsev}$^{2}$}

\author{Philippe \surname{Grangier}$^{2}$}

\author{Anders~S. \surname{S\o rensen}$^{1}$}

\affiliation{$^1$ The Niels Bohr Institute, University of Copenhagen, Blegdamsvej 17, DK-2100 Copenhagen  \O, Denmark\\
$^2$ Laboratoire Charles Fabry, Institut d'Optique, CNRS, Univ Paris-Sud, Campus Polytechnique, RD 128, 91127 Palaiseau cedex, France\\
$^3$ Laboratoire Aim\'{e} Cotton, CNRS, Universit\'{e} Paris Sud, ENS Cachan, 91405 Orsay, France.\\
$^4$ Department of Physics, Harvard University, Cambridge, MA 02138, USA}
\date{\today}

%%%%%%%%%%%%%%%%%%%%%%%%%%%%%%%%%%%%%%%%%%%%%%%%%%%%%%%%%%%%%%%%%%%%%%%%%%%%%%%%
%%%%%%%%%%%%%%%%%%%%%%%%%%%%%%%%%%%%%%%%%%%%%%%%%%%%%%%%%%%%%%%%%%%%%%%%%%%%%%%%

\begin{abstract}
We propose a novel scheme for high fidelity photonic controlled-phase gates using Rydberg blockade in an ensemble of atoms  in an optical cavity.
The gate operation is obtained by first storing a  photonic pulse in the ensemble and then scattering a second pulse from the cavity, resulting in a phase change depending on whether the first pulse contained a single photon. We show that the combination of Rydberg blockade and optical cavities effectively enhances the optical non-linearity created by the strong Rydberg interaction and makes the gate operation more robust. The resulting gate can be implemented with cavities of moderate finesse allowing for highly efficient processing of quantum information encoded in photons. As an illustration, we show how the gate can be employed to increase the communication rate of quantum repeaters based on atomic ensembles. 
\end{abstract}

%%%%%%%%%%%%%%%%%%%%%%%%%%%%%%%%%%%%%%%%%%%%%%%%%%%%%%%%%%%%%%%%%%%%%%%%%%%%%%%%
%%%%%%%%%%%%%%%%%%%%%%%%%%%%%%%%%%%%%%%%%%%%%%%%%%%%%%%%%%%%%%%%%%%%%%%%%%%%%%%%
\maketitle
%%%%%%%%%%%%%%%%%%%%%%%%%%%%%%%%%%%%%%%%%%%%%%%%%%%%%%%%%%%%%%%%%%%%%%%%%%%%%%%%
%-----------Text body-----------------------------------------------------------

Large bandwidth, fast propagation and the non-interacting nature of photons, make them ideal for communicating quantum information over long distances \cite{Kim08}. In contrast, strong photon-photon interactions are desirable for processing of quantum information encoded in the photons, especially if both high fidelity and high efficiency are needed. To satisfy these requirements one needs a highly non-linear medium. Typically, the strength of photon-photon interactions mediated by a non-linear medium is very weak at the single-photon level where photonic quantum logic gates are operating \cite{Chang14}. As a consequence, the implementation of photonic quantum gates remains an unsolved challenge and requires novel means of efficient light-matter interaction. To enhance light-matter interactions, a viable solution is to use ensembles of atoms, e.g., configured for electromagnetically induced transparency (EIT) \cite{AS10}; this can be further improved by placing the ensemble in an optical cavity, but these ensemble based approaches do not increase the essential photonic non-linearity. In recent years, there has been intense efforts to realize light-matter interactions via, non-linear interactions in a variety of medium, ranging from atoms \cite{Turch95, Har99, Dar05, Tey08,Lukin13,Tob14, Andre14} and atom like systems \cite{Mich00, Fush08, Hwang09, Casa15} to superconducting qubits \cite{Dev13, Pra13, Neu13}.

A promising approach towards creating strong quantum nonlinearities is to exploit  excitation blockade in Rydberg EIT systems \cite{Frie05, Pri10, Gor11, Pet11, Max13, Stano13, Pey13, Bau14}. Several quantum effects like strong optical non-linearities and control of light by light \cite{Pri10,Pey13,Max13,Bau14,Parigi12, Hoff14,Tiark14}, deterministic single-photon sources \cite{Dud12}, and the generation of entanglement and atomic quantum gates \cite{Gran09, Mul09, Li13, Adam14, Khaz15} have been investigated. The strong nonlinearity originates from the fact that the Rydberg  interaction prevents multiple excitations within a blockaded radius $r_{b}$ \cite{Luk01, Saf10}. The ensemble then behaves as a two-level \textit{superatom} consisting of $N_b$ atoms within a radius $r_{b}$ \cite{Luk01, Saf10}. If the optical depth $d_{b}$ corresponding to the superatom is sufficiently large, $d_b \gg 1$ \cite{Pri10}, a strong optical nonlinearity at the single-photon level can be achieved in the EIT configuration \cite{Frie05, Pri10, Gor11, Pet11, Max13, Stano13, Pey13, Bau14}. Reaching such an optical depth is, however, challenging, which limits the effectiveness of photonic quantum gates. 

An enhanced optical nonlinearity was recently demonstrated by placing the ensemble inside an optical cavity \cite{Parigi12}, but a direct application of this nonlinearity for quantum gates is non-trivial since the outgoing optical modes are highly distorted and entangled by the interaction \cite{Shapiro06, Gea10, Bing12}. In this letter, we propose a novel scheme for achieving a high fidelity \textit{photonic controlled-phase (CP) gate with a Rydberg EIT ensemble} trapped inside an optical cavity of moderate finesse. In our scheme, the photons are incident at different times thus avoiding the problem of mode distortion while still allowing the cavity enhancement of the interaction. The use of a cavity has several major advantages compared to ensembles in free space, since it enhanced light-atom coupling in the ensemble and also effectively increases the non-linearity. In our proposal, the parameter characterizing the Rydberg blockade is $\mathcal{C}_b \sim \mathcal{F} d_b$, where $\mathcal{C}_{b} = N_{b}\mathcal{C}$, with $\mathcal{C} \ll 1$ being the single atom cooperativity, and $\mathcal{F}$ is the cavity finesse. Hence the effect of the Rydberg interaction is increased by the cavity finesse $\mathcal{F}$, whereas the low value of $\mathcal{C}$ is compensated by a high value of $N_b$. In addition, the cavity is also useful for controlling the mode structure thereby enabling high input-output efficiencies \cite{Grang14}. We show that the proposed gate can have a promising (heralded) error scaling as $1/C_b^2$, and demonstrate how it can be used to improve quantum repeaters based on atomic ensembles even for moderate interactions strengths $C_b\sim 10$. The proposed CP gate can thus be directly integrated into quantum communication circuitry thereby providing a building block for future quantum networks. The Rydberg interaction \cite{Saf10} for our proposal can either be long range dipolar or van der Waals interactions, but for simplicity, we only consider the latter.

We first outline the basic idea of our gate, which goes along the line of Ref. \cite{Duan04}, except that the single trapped atom is replaced by a Rydberg ensemble. In contrast to Ref. \cite{Duan04}, and many others, we thus do \textit{not} require the strong-coupling regime of cavity QED, and can  work with cavities of moderate finesse. This enables input-output efficiency near unity since  the cavity losses can be completely negligible compared to the mirror's transmission \cite{Grang14}. For simplicity, we first describe the operation for single-rail qubits where a qubit is encoded in a photon pulse containing a superposition of vacuum $|0\rangle$ and a single photon $|1\rangle$. Later, we generalize it to a more useful dual-rail encoding where the qubit is encoded as a photon in one of two possible modes. 

\begin{figure}[!h]
   \begin{center}
   \begin{tabular}{c}
  \includegraphics[height=6.5cm]{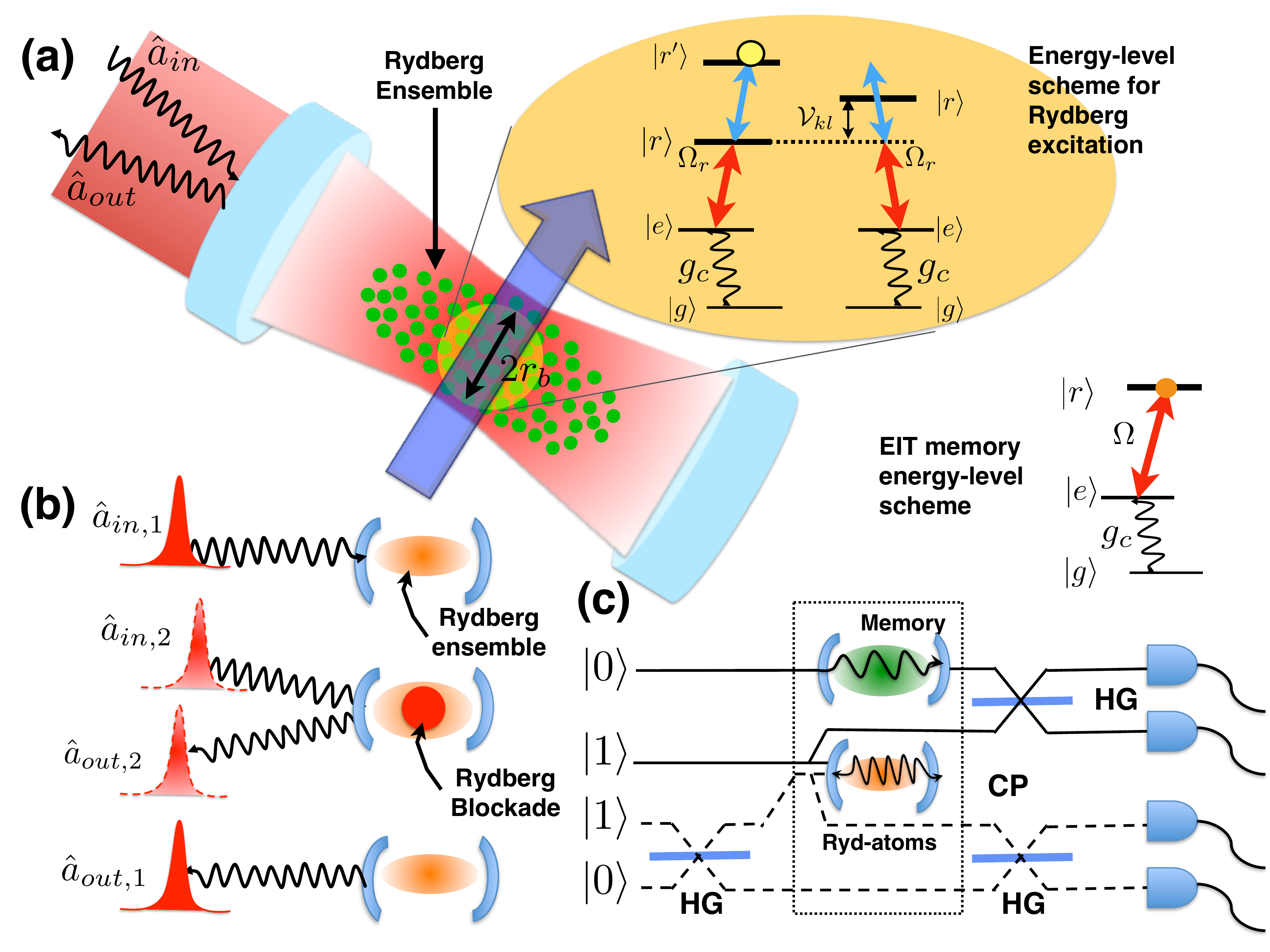}
   \end{tabular}
   \end{center}
   \caption[example] 
   { \label{fig1} 
Schematic outline of the phase gate (a) An input single photon pulse along with a driving field induces a two-photon transition to the Rydberg state $|r\rangle$ which is subsequently transferred to another Rydberg state $|r'\rangle$. Due to Rydberg  interactions $\mathcal{V}_{kl}$ among the atoms, other Rydberg states $|r\rangle$ within the range of the interaction potential, given by, the blockade radius of $r_{b}$, 
become off-resonant allowing no further excitation. (b) When an initial photon pulse is stored in the Rydberg ensemble, the second incoming photon cannot enter the cavity 
and is scattered off, which ideally induces a phase flip of $\pi$ on the scattered photons. (c) Dual-rail implementation of a CP gate (dotted box). A Bell state measurement can be implemented by combining the CP gate with Hadamard gates (HG).}
\end{figure}

In the single-rail version outlined in Figs. \ref{fig1}(a) and (b), a first photon pulse is stored in a cavity containing a Rydberg EIT ensemble \cite{Gor7}. Here a classical driving field from an excited state $|e\rangle$ to a Rydberg state $|r\rangle$ enables the storage of incoming photons in $|r\rangle$ through the interaction of the cavity field with the transition from the ground state $|g\rangle$ to $|e\rangle$. The excitation in state $|r\rangle$ is then transferred to another Rydberg state $|r'\rangle$ by a microwave pulse so that the ensemble contains a single atom in state $|r'\rangle$ if the first pulse contained a single incoming photon. The second pulse is then incident on the cavity. If the first pulse contained vacuum $|\O\rangle$, the second pulse is scattered under Rydberg EIT conditions and leaves the cavity with the same phase. If the first pulse contained a photon, the atom in $|r'\rangle$ shifts the position of the state $|r\rangle$ in the remaining atoms. As we will show, this prevents the second pulse from entering the cavity resulting in a phase flip on the $|1\rangle$ component of the second pulse. This evolution thus realizes a CP gate which, together with single qubit operations, is universal for quantum information processing. 
%%%--------------------------------------------------------------------------
%%%--------------------------------------------------------------------------  

We now present a theoretical treatment to evaluate the performance of the CP gate. The initial state of the single photon pulse can be expressed as $\int d\omega \phi(\omega)\hat{a}^{\dagger}_{\omega}e^{-i\omega t}|\O\rangle$, where $\phi(\omega)$ is the normalized pulse shape, $\hat{a}^{\dagger}_{\omega}$ is the one-dimensional field operators satisfying the standard bosonic commutation relations and $|\O\rangle$ denotes the vacuum of all the optical modes. The frequency integrand $\omega$ of the incoming photon is referenced to the cavity frequency $\omega_{c}$, which in turn is nearly resonant to the $|e\rangle \rightarrow |g\rangle$ transition (see Fig. \ref{fig1}(a)). The cavity is assumed to be one-sided with a standing-wave field. The dynamics of the system can be described in the quantum jump approach through the no-jump Hamiltonian $\mathcal{H} = \mathcal{H}_{s}+\mathcal{H}_{I}$. Here $\mathcal{H}_{s}$ consists of the decays and the free energy terms \cite{supp} while,  
\bea
\mathcal{H}_{I} & = & -\sum_{l}\hbar\left[\frac{\Omega_{l}}{2}|r_{l}\rangle\langle e_{l}|+i\mathcal{G}_{l}|e_{l}\rangle\langle g_{l}|\hat{b}\right]+\text{H.c.}\nonumber\\
& &+\sum_{k}\hbar\mathcal{V}_{kl}|r'_{k}\rangle\langle r'_{k}|\otimes|r_{l}\rangle\langle r_{l}|,
\eea
Here the coupling strengths of the $l^\text{th}$ atom with the driving field and the incoming single photon pulse is respectively $\Omega_{l}$ and $\mathcal{G}_{l}$, while $\mathcal{V}_{kl}$ is the van der Waals interaction among the Rydberg excitations of atoms  $k$ and $l$.
We solve the Schr\"{o}dinger equation for the scattering stage assuming constant  $ \Omega_l$ in Fourier space to find the reflection co-efficient. 
The (amplitude) reflection coefficient with the stored Rydberg excitation in atom $k$ is given by $\mathcal{R}_{k}(\omega) = \left(2\kappa\mathbf{S}_{k}(\omega)-1\right)$ where,
\be
\label{3}
\mathbf{S}_{k}(\omega)=\left(\kappa-i\omega+\sum_{l}\frac{|\mathcal{G}_{l}|^2}{(\Gamma_{el}-i\tilde\Delta_{l})+\frac{|\Omega_{l}/2|^2}{\Gamma_{rl}+i(\delta_{l}+\mathcal{V}_{kl}-\omega)}}\right)^{-1}.
\ee
where the detunings are $\Delta_{l} = \omega_{e_{l}}-\omega_{c}$, $\tilde{\Delta}_{l} = \Delta_{l}+\omega$ and $\delta_{l} = (\omega_{r_{l}}-\omega_{e})-\omega_{c}$ with $\hbar \omega_{e_l(r_l)}$ and $\Gamma_{el(rl)}$ being the energy and width of the excited (Rydberg) state $|e\rangle (|r\rangle)$ in atom $l$ and $\kappa$ is the cavity field decay rate.  The reflection coefficient $\mathcal{R}_{g}(\omega)$  \cite{supp} for no stored excitation is evaluated by setting $\mathcal{V}_{kl} = 0$.

To get an understanding of the scattering we study the behavior  of the reflection coefficient for resonant interactions $\delta_{l} = \Delta_{l} = 0$ and long lived   Rydberg excitations $(\Gamma_{rl} = 0)$.  For simplicity, we assume equal couplings and driving strengths on all atoms $\mathcal{G}_{l} = \mathcal{G}$ and $\Omega_l=\Omega$ (for the general case see \cite{supp}). Furthermore, if the photon pulse has a suitably long duration we can put $\omega \approx 0$ (see below). With these assumptions, we find from Eq. (\ref{3}) that $\mathcal{R}_{g} = 1$ for no stored excitation; this is the perfect EIT condition. When an excitation is stored, the reflection coefficient becomes
\be
\label{5}
\mathcal{R}_{k} = \left(\frac{2}{1+\mathcal{C}^{\ast}_{v}}-1\right),
\ee
where $\mathcal{C}^\ast_{v} =\mathcal{C}_{b}+i\mathcal{C}'_{b} = \mathcal{C} \sum_{l}1/\left[1+\frac{|\Omega/2|^{4}}{\mathcal{V}^{2}_{kl}\Gamma^{2}_{e}}\right]+i \mathcal{C} \sum_{l}\frac{|\Omega/2|^{2}}{\mathcal{V}_{kl}\Gamma_{e}}/\left[1+\frac{|\Omega/2|^{4}}{\mathcal{V}^{2}_{kl}\Gamma^{2}_{e}}\right]$ quantifies the effective cooperativity of the blockade, while $\mathcal{C} = |\mathcal{G}|^{2}/(\kappa\Gamma_{e})$ is the single atom cooperativity. Each atom $l$ in the volume blocked by the $|r'_k\rangle$ excitation, \emph{i.e.} such that $\mathcal{V}_{kl} \gg |\Omega/2|^{2}/\Gamma_{e}$, will contribute with $\mathcal{C}$ in $\mathcal{C}_b$. On the other hand, those atoms for which $\mathcal{V}_{kl} \ll |\Omega/2|^{2}/\Gamma_{e}$ will have a negligible contribution to $\mathcal{C}_b$, and hence $\mathcal{C}_b$ gives the effective cooperativity of the blockade ensemble. The imaginary part depends on the shape of the interaction but for a uniform $1/r^6$ interaction at resonance in a uniform cloud we find that $|\mathcal{C}'_b| = \mathcal{C}_b$ \cite{supp} ($r$ is the distance between atoms $k$ and $l$). 

We now discuss the key feature of our work - the implementation of a photonic CP gate via scattering from a Rydberg ensemble in either a single-rail or  dual-rail encoding. The  single-rail implementation uses  the encoding discussed in the introduction and is shown schematically in Fig.  \ref{fig1} (b).  A first qubit is encoded in the vacuum and single photon state, $|\O\rangle$  and  $|1\rangle = \int d\omega \phi(\omega) \hat{a}^\dagger_{\omega}e^{-i\omega t}|\O\rangle$, respectively, of a first incoming pulse. This pulse is stored in the Rydberg ensemble such that the logical states $|0\rangle$ and $|1\rangle$ are mapped onto the ensemble being in the joint ground state $|0\rangle=|g^N\rangle|\O\rangle$ and a Rydberg polariton $|1\rangle = \sum_k \alpha_{k}|g^{N-1}, r'_k\rangle|\O\rangle$. This is achieved using the well established techniques of storage in atomic ensembles, which is known to have an error $1/N\mathcal{C}$ for any slowly varying pulse shape provided a temporally varying control field is used during storage \cite{AS10,Gor11}, followed by microwave $\pi$-pulse between $|r_k\rangle$ and $|r_k'\rangle$. A second incoming photon pulse is then reflected from the cavity. This reflection can be from either an ensemble in the EIT configuration (ensemble in $|0\rangle$), or from a blocked ensemble ($|1\rangle$). As can be seen from Eq. (\ref{5}) there is exactly a $\pi$ phase shift between the two situations: $\mathcal{R}_g=1$ for $\mathcal{C}^*_v=0$ and $\mathcal{R}_k=-1$ for $|\mathcal{C}^*_v|\gg 1$. Finally, the first stored pulse is retrieved from the ensemble. 

To evaluate the performance we calculate the Choi-Jamiolkowski fidelity of the gate. Since, in general, we have $N\mathcal{C} > \mathcal{C}_b$, the fidelity of the operation will mainly be limited by the gate and we shall ignore imperfections during the storage. The fidelity can then be determined by \cite{supp} ,
\bea
\label{7}
F_\text{CJ}&=&\frac{1}{16}\Big|2+\int d\omega |\phi(\omega)|^{2}\mathcal{R}_{g}(\omega)\nonumber\\
&-&\sum_{k}\int d\omega|\alpha_{k}|^{2}|\phi(\omega)|^{2}\mathcal{R}_{k}(\omega)\Big|^{2}.
\eea 
To account for errors due to imperfect Rydberg blockade, we evaluate the above fidelity and find
\bea
\label{8}
F_\text{CJ} & = & 1-\frac{(1+\mathcal{C}_{b})}{(1+\mathcal{C}_{b})^{2}+\mathcal{C}^{'2}_{b}}-\frac{N\mathcal{C}\Gamma^{2}_{e}}{|\Omega/2|^{4}}(\Delta\omega)^{2}\nonumber\\
& &-\left(\frac{1}{\kappa}+\frac{N\mathcal{C}\Gamma_{e}}{|\Omega/2|^{2}}\right)^{2}(\Delta\omega)^{2}
\eea
Here the third and fourth term are gate errors due to the finite frequency width $\Delta\omega^2$ of the incoming pulse. These terms arise predominately from the EIT bandwidth, which is much narrower than the variations of the blocked reflection coefficient. For a narrow pulse $\Delta\omega\rightarrow 0$, the fidelity is only limited by the  cooperativity of the blocked ensemble $1-F_\text{CJ}\propto1/\mathcal{C}_b$. Hence, as discussed in the introduction, it is the cavity enhanced blockaded cooperativity, which is the main figure of merit for the gate. 

\begin{figure}[!h]
   \begin{center}
   \begin{tabular}{c}
   \includegraphics[height=4cm]{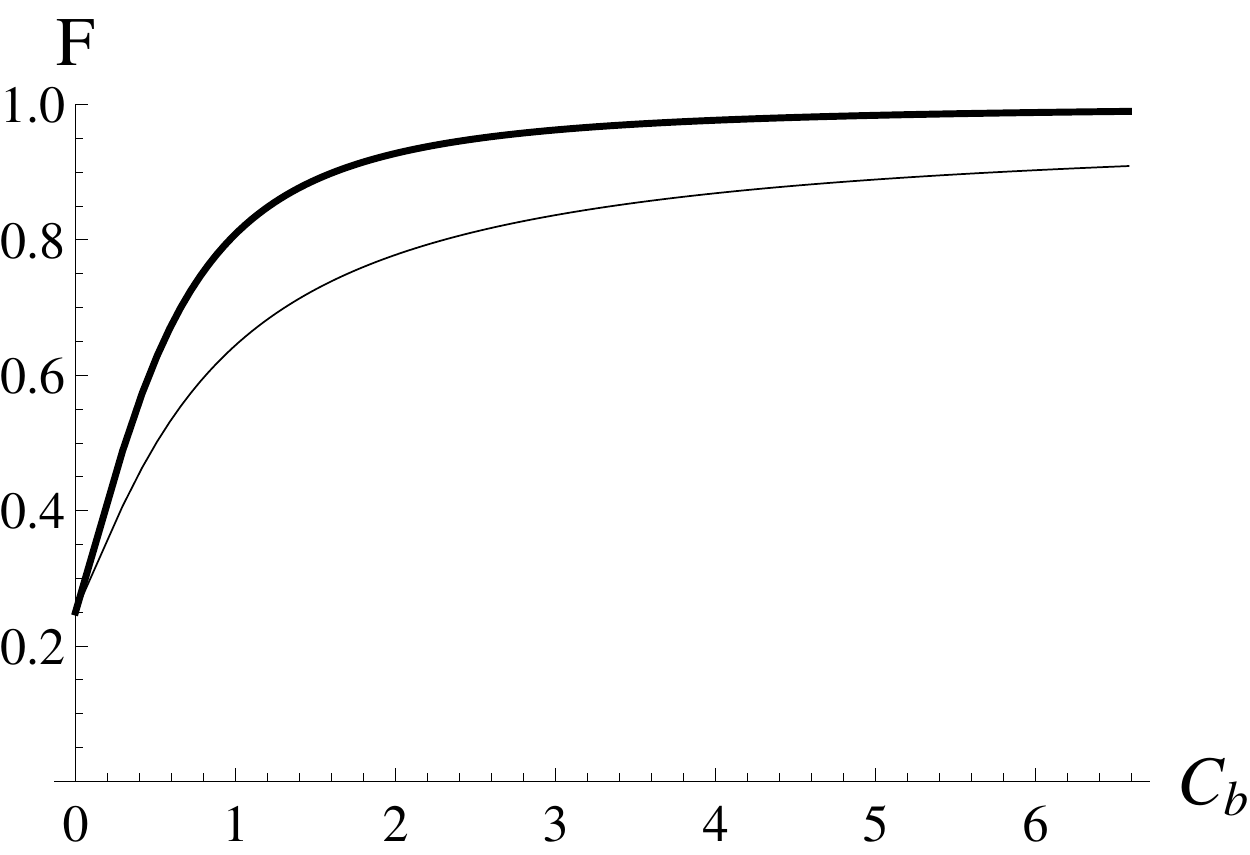}
   \end{tabular}
   \end{center}
   \caption[example] 
   { \label{plot} Choi-Jamiolkowski fidelity (thin line) and post-selected swap
fidelity (thick line) as functions of the blockaded cooperativity $\mathcal{C}_b$ for a spectrally narrow pulse $\Delta\omega\rightarrow 0$. 
We assume $|\mathcal{C}'_b|=\mathcal{C}_b$.}
\end{figure}

In the dual-rail encoding, both logical states $|0\rangle$ and $|1\rangle$ are represented by photons, but in two different paths. A schematic of the dual-rail CP gate is shown in Fig. \ref{fig1}(c). The first photon pulse in the upper two arms of the figure is first stored in a memory consisting of a Rydberg ensemble placed in each arm (for a polarization encoding such two memories might be realized by two different internal states of the same ensemble). A second photon pulse is then scattered from the Rydberg ensemble if it is in state $|1\rangle$ (upper rail in the figure). This scattering ideally induces a phase change of $\pi$  if there was a photon stored in the Rydberg ensemble, i.e. if both qubits were in state $|1\rangle$. As opposed to the single-rail implementation, the dual-rail implementation has the possibility of conditioning on getting two photons in the output. Since the dominant error in the single-rail implementation is the loss of photons,  this possibility  allows for a substantial increase in the fidelity with only a minor failure probability of the gate. In view of a possible application of the gate for quantum repeaters, discussed below, we  consider the  conditional fidelity of an EPR pair resulting from an entanglement swap realized with the gate using the full circuit in Fig. \ref{fig1}(c). Neglecting again the error due to finite storage efficiency, we find  that this fidelity is \cite{supp}
\bea
\label{9}
F_\text{swap} = \frac{\int d\omega|\phi(\omega)|^{2}\left|2+\mathcal{R}_{g}(\omega)-\sum_{k}|c_{k}|^{2}\mathcal{R}_{k}(\omega)\right|^{2}}{16P_\text{suc}}
\eea
where the success probability of the process is $P_\text{suc} = \int~d\omega|\phi(\omega)|^{2}(2+ |\mathcal{R}_{g}(\omega)|^{2}+|\sum_{k}|c_{k}|^{2}\mathcal{R}_{k}(\omega)|^{2})/4$. Note that compared to $F_\text{CJ}$ in Eq. (\ref{7}), the only difference is due to the conditioning with a success probability $P_\text{suc}<1$ and the way the mode function is treated. The latter is related to the fact that  Eq. (\ref{7}) is the fidelity with a specific mode function. Keeping only the leading order contribution to the dispersion, we find the fidelity and success probability of the CP gate 
\bea
\label{10}
F_\text{swap} & = & 1-\frac{1}{[\mathcal{C}^{2}_{b}+\mathcal{C}^{'2}_{b}]}-\frac{3\mathcal{C}^{2}_{b}-\mathcal{C}^{'2}_{b}}{4[\mathcal{C}^{2}_{b}+\mathcal{C}^{'2}_{b}]^{2}}\nonumber\\
&&-\frac{3}{4}\left[\frac{1}{\kappa}+\frac{N\mathcal{C}\Gamma_{e}}{|\Omega/2|^{2}}\right]^{2}(\Delta\omega)^{2},\\
\label{11}
P_\text{suc} & = & 1-\frac{\mathcal{C}_{b}}{(1+\mathcal{C}_{b})^{2}+\mathcal{C}^{'2}_{b}}-\frac{N\mathcal{C}\Gamma^{2}_{e}}{|\Omega/2|^{4}}(\Delta\omega)^{2}
\eea
Here the fourth term is again the leading order error from the spectral width of the pulse. In the limit of a narrow pulse $\Delta\omega\rightarrow 0$, we see that the conditional gate error  $1-F_\text{swap}\propto 1/(\mathcal{C}_b)^2$ for $\mathcal{C}_b\gg 1$ is much smaller than for the single-rail. This comes at only a minor cost in the failure probability $1-P_\text{suc}\propto 1/\mathcal{C}_b$. The resulting dual-rail fidelities are plotted in Fig. \ref{plot} as a function of the parameter $\mathcal{C}_{b}$. For $\mathcal{C}_{b}\approx 8$, the (post-selected) fidelity is found to be larger than $0.99$

In order to get realistic predictions, we use the experimental conditions of Ref. \cite{Parigi12}, with $\Gamma_{e}\approx(2\pi)3\mbox{MHz}$ and $\kappa\approx(2\pi)10\mbox{MHz}$ (corresponding to a finesse $\mathcal{F}\approx120$) but a smaller beam waist $w_{0}=15\mu\mathrm{m}$. This gives a single atom cooperativity ${\cal C}=0.025$ and we take $N{\cal C}= 20$ corresponding to a combined storage and retrieval efficiency of 90\%. We assume a Rydberg line width $\gamma_r=(2\pi) 60$ kHz \cite{low} corresponding to a coherence time of $\tau_r=1/\gamma_r = 2.65\ \mu$s (note that if the two ensembles in the dual-rail encoding are read out with the same laser the scheme becomes insensitive to phase fluctuations). With a pulse duration of $T=1/\Delta\omega=300$ ns and a driving strength of $\Omega=(2\pi)36$ MHz the error due to finite bandwidth in Eq. (\ref{10}) is below 2\%. Taking the interaction  $\mathcal{V}= (2\pi)8.31\cdot10^{6}/r^6~\mbox{MHz}~\mu\mbox{m}^{6}$ corresponding to two atoms with a Rydberg quantum number $n_{r} =  90$ and an atomic density of $n=0.25\ \mu\mbox{m}^{-3}$, one has $\mathcal{C}_{b}\sim 8.1$ \cite{supp} which is sufficient to obtain high fidelities as show in Fig. \ref{plot}. Here, we ignore any effect of sample inhomogeneities, but this can be taken into account by suitable redefinitions of $\mathcal{C}_b$ and $\mathcal{C}'_b$ \cite{supp}.

\begin{figure} 
\centering
\includegraphics[height = 4.5cm]{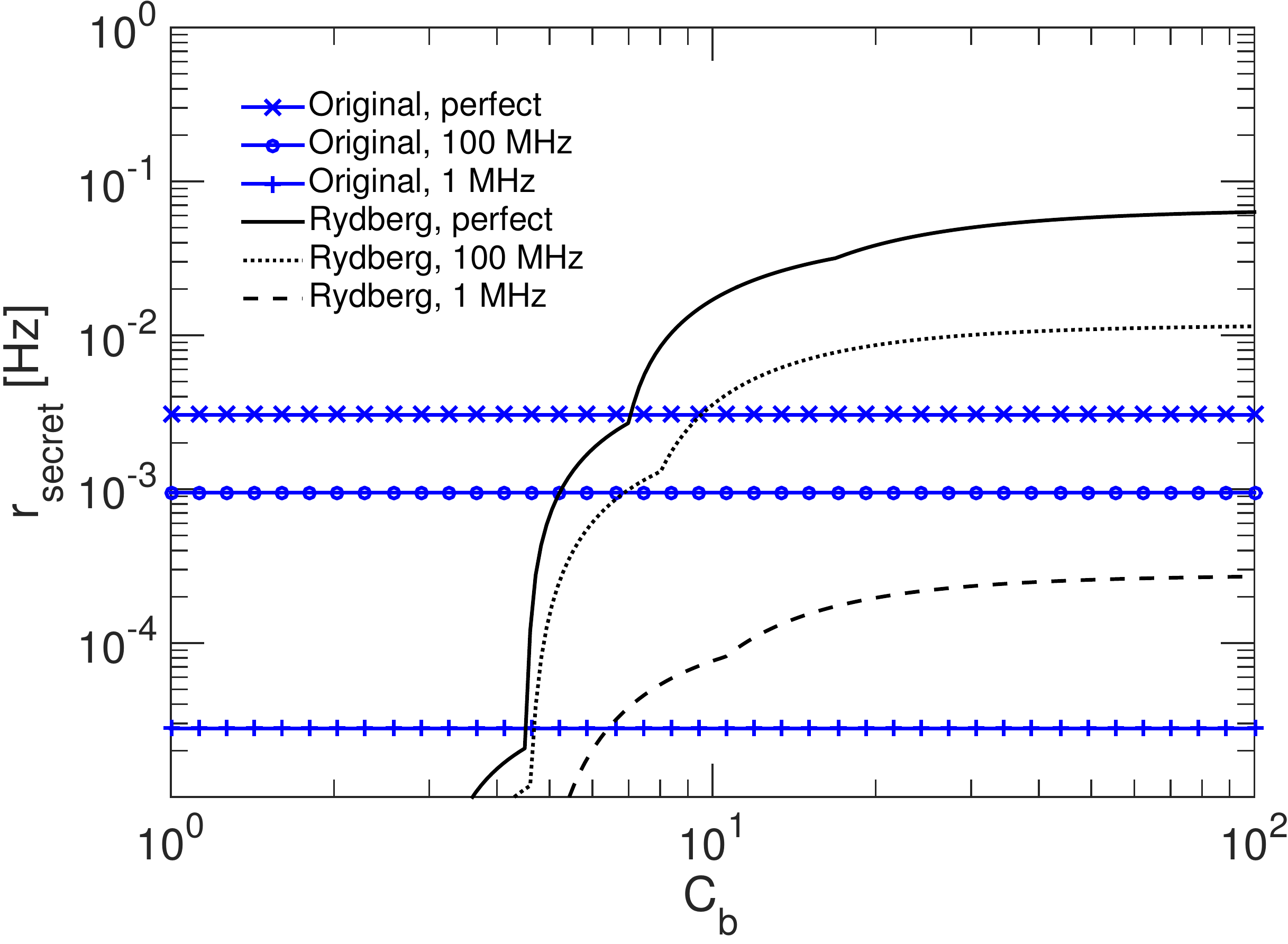}
\caption{Secret key rate per repeater station ($r_{\text{secret}}$) as a function of the blockaded cooperativity ($\mathcal{C}_{b}$) for a communication distance of 1000 km. We  compare the protocol of Ref.~\cite{sang} (Original) with a modified protocol where the entanglement swapping is performed with the Rydberg CP gate (Rydberg). We consider an optimistic source repetition rate of 100 MHz and a more modest one of 1 MHz, as well as a perfect single excitation state created in the atomic ensembles, e.g., using Rydberg blockade \cite{Dud12}. We assume an attenuation length of 22 km in the fibers and an optical signal speed of $2\cdot10^{5}$ km/s. The ensemble readout efficiency and photodetector efficiency are both assumed to be 90\%. The steps in the curves reflect where the fidelity of the CP gate allows additional swap levels to be employed.  }
\label{fig:repeater}
\end{figure}

As a particular application of the gate, we consider long distance quantum cryptography based on quantum repeaters. We considered the ensemble based quantum repeater protocol proposed in Ref.~\cite{sang}, 
but replace the entanglement swapping with the procedure shown in Fig. \ref{fig1}(c). We calculate the secret key rate per repeater station as described in Ref.~\cite{JB2} (assuming the distributed states to be Werner states) and compare the results to the original protocol (see Fig.~\ref{fig:repeater}). At the lowest level of the protocol, single excitations are stored in atomic ensembles using a Raman scheme and we include double excitation errors to lowest order similar to Ref.~\cite{sang}.  The performance of the protocol depends strongly on the repetition rate of this operation. Regardless of the repetition rate, the CP gate enables significantly higher communication rates since it allows near perfect Bell state measurements (for $\mathcal{C}_{b}\gg1$) whereas swapping operations based on linear optics have a maximal success probability of 50\%. In Fig.~\ref{fig:repeater}, we also show the rate obtainable if single excitations are initially created perfectly and deterministically in the atomic ensembles, e.g., by exploiting Rydberg blockade \cite{Dud12}. We find that for such a protocol, a cooperativity of $\mathcal{C}_{b}\sim25$ is sufficient to obtain a secret key rate of 1.5 Hz over 1000 km using 33 repeater stations. 

In conclusion, we have proposed an efficient method to implement a CP gate for photons. The gate combines the advantages of cavity defined optical modes and cavity enhanced light matter interactions with the strong Rydberg blockade obtainable in atomic ensembles. As a direct application, the proposed gate can be used to improve the communication rate of quantum repeaters, but more generally the gate may serve as a building block for photonic quantum networks.

\begin{acknowledgments}
The research leading to these results was funded by the European Union Seventh Framework Programme through SIQS (Grant No. 600645) and ERC Grant QIOS (Grant No. 306576). J.B. acknowledges funding from the Carlsberg foundation. S.D. and A.G. contributed equally to this work.
\end{acknowledgments}

\newpage

\renewcommand{\theequation}{S\arabic{equation}}
\renewcommand{\thesection}{S\arabic{section}}
\renewcommand{\thefigure}{S\arabic{figure}}

\onecolumngrid

\begin{center}
\Large{Supplementary Material}
\end{center}

\section{The Reflection Coefficient}\label{sec_reflection_coefficient}
\noindent{}The dynamics of the Rydberg ensemble in the cavity can be described through the no-jump Hamiltonian $\mathcal{H}$ consisting of the free energy and decay terms $\mathcal{H}_\text{s}$ along with the interaction part as $\mathcal{H}_{I} = \mathcal{H}_\text{L-int}+\mathcal{H}_\text{Ryd-int}$, where 
\bea
\label{1}
&&\mathcal{H}_\text{s} =\sum_{l}\hbar(\Delta_{l}-i\Gamma_{el})|e_{l}\rangle\langle e_{l}|+\hbar(\delta_{l}-i\Gamma_{rl})|r_{l}\rangle\langle r_{l}|-i\hbar\kappa\hat{b}^\dagger \hat{b}\nonumber\\
&&\mathcal{H}_\text{L-int} = -\sum_{l}\frac{\hbar\Omega_{l}}{2}|r_{l}\rangle\langle e_{l}|-i\sum_{l}\hbar\mathcal{G}_{l}|e_{l}\rangle\langle g_{l}|\hat{b}+\text{H.c.}\nonumber\\
&&\mathcal{H}_\text{Ryd-int} = \sum_{k}\hbar\mathcal{V}_{kl}|r'_{k}\rangle\langle r'_{k}|\otimes|r_{l}\rangle\langle r_{l}|,
\eea
where the detunings $\Delta_{l},\delta_{l}$, the linewidths $\Gamma_{e}, \Gamma_{r}$, and the coupling strengths $\Omega_{l}, \mathcal{G}_{l}$ are as defined in the main text, while $2\kappa$ is the cavity intensity decay rate. Note that all energies are measured relative to the cavity resonance, and hence the cavity term in $\mathcal{H}_\text{s}$ only involve the loss rate $\kappa$.  are as defined in the main text, while $\mathcal{V}_{kl}$ is the van der Waals interaction among the Rydberg excitations of atoms  $k$ and $l$. The incoming and outgoing photons are going to be accounted for by the input-output relations. After the storage of the first pulse, the wave-function of the combined field and ensemble with the initial excitation stored in $|r'_{k}\rangle$ and one incoming photon is given by, 
\bea
\label{2}
|\Psi\rangle  & = & \sum_{k}\int d\omega\beta_k(\omega)\hat{a}^{\dagger}_{\omega}e^{-i\omega t}|g^{N-1},r'_{k},\O\rangle+\sum_{k} \hat{b}^{\dagger}C_{bk}|g^{N-1},r'_{k},\O\rangle\nonumber\\
&+&\sum_{l}\Big\{C_{ekl}|g^{N-2},e_{l},r'_{k},\O\rangle+C_{rkl}|g^{N-2},r_{l},r'_{k},\O\rangle\Big\}.
\eea
Here $|\O\rangle$ is the vacuum state, $C_{ekl}$ and $C_{rkl}$ are respectively the amplitude of being in the excited state $|e\rangle$ and the Rydberg state $|r\rangle$ when there is one stored Rydberg excitation in the $k^\text{th}$ atom, while $C_{bk}$ is the amplitude of the cavity excited state. We next evaluate the Schr\"odinger 
equation for the wave-function (\ref{2}) together with the input-output relations to find the dynamical behavior of the amplitudes $C_{ak},C_{ekl},C_{rkl}$
\bea
\label{eq.ca}
\dot{C}_{bk} & = & \sum_{l}C_{ekl}\mathcal{G}^{\ast}_{l}-\kappa C_{bk}+\sqrt{2\kappa}\beta^\text{in}_{k},\\
\label{eq.cl}
\dot{C}_{ekl} & = & -i(\Delta_{l}-i\Gamma_{el})C_{ekl}+i\frac{\Omega_{l}^{\ast}}{2}C_{rkl}-\mathcal{G}_{l}C_{bk},\\
\label{eq.cll}
\dot{C}_{rkl} & = & -i(\delta_{l}-i\Gamma_{rkl})C_{rkl}+i\frac{\Omega_{l}}{2}C_{ekl}-iC_{rkl}\mathcal{V}_{kl}.
\eea
The outgoing field amplitude is then given by,
\be
\label{3}
\beta^\text{out}_{k}(\omega) = \sqrt{2\kappa}C_{bk}(\omega)-\beta^\text{in}_{k}(\omega), 
\ee
where $C_{bk}$ is found by solving the set of Eqns. (\ref{eq.ca}-\ref{eq.cll}) using Fourier transformation. We thereby get, 
\bea
\label{4}
C_{bk}(\omega) & = & \sqrt{2\kappa}\beta^\text{in}_{k}(\omega)\mathbf{S}_{k}(\omega)\nonumber\\
\mathbf{S}_{k}(\omega)& = &\left(\kappa-i\omega+\sum_{l}\frac{|\mathcal{G}_{l}|^2}{(\Gamma_{el}-i\Delta_{l}-i\omega)+\frac{|\Omega_{l}/2|^2}{\Gamma_{rl}+i(\delta_{l}+\mathcal{V}_{kl}-\omega)}}\right)^{-1}.
\eea
Substituting $C_{bk}$ into Eq. (\ref{3}) we get,
\be
\label{4a} 
\mathcal{R}_{k}(\omega) = 2\kappa\left(\kappa-i\omega+\sum_{l}\frac{|\mathcal{G}_{l}|^2}{(\Gamma_{el}-i\Delta_{l}-i\omega)+\frac{|\Omega_{l}/2|^2}{\Gamma_{rl}+i(\delta_{l}+\mathcal{V}_{kl}-\omega)}}\right)^{-1}-1.
\ee
Assuming all fields to be resonant i.e. for  $\delta_{l} = \Delta_{l} = 0$, a long lived Rydberg state ($\Gamma_{rl} = 0$), and slowly varying photon pulses $(\omega = 0)$ we get,
\be
\label{5}
\mathcal{R}_{k}  = 2\left(1+\sum_{l}\frac{|\mathcal{G}_{l}|^{2}/\kappa\Gamma_{e}}{1-i|\Omega_{l}/2|^{2}/\mathcal{V}_{kl}\Gamma_{e}} \right)^{-1}-1.
\ee
which under the assumption of equal coupling strengths $\mathcal{G}_{l} = \mathcal{G}$ and Rabi frequencies $\Omega_{l}= \Omega$, for the defined single atom co-operativity $\mathcal{C} = |\mathcal{G}|^{2}/\kappa\Gamma_{e}$ becomes, $\mathcal{R}_{k} = \left[2\left(1+\mathcal{C}^{\ast}_{v}\right)^{-1}-1\right]$, where $\mathcal{C}^\ast_{v} = \sum_{l}\mathcal{C}/\left(1-i|\Omega/2|^{2}/\mathcal{V}_{kl}\Gamma_{e}\right)$. To get a simple physical understanding of the scattering dynamics, we shall first assume that all atoms are identical (homogeneous). We will consider what happens for an inhomogeneous ensemble in a later section. For the van der Waals interaction potential $\mathcal{V}_{kl} = -C_{6}/r^{6}$, where $r$ is the relative distance between the $k^{th}$ and $l^{th}$ atoms, we can evaluate $\mathcal{C}^\ast_{v} $ with the sum $\sum_{l}$ converted to a volume integral $\rightarrow \int n dV$. Thus we get for a homogeneous ensemble with an isotropic potential,
\be
\label{6}
\mathcal{C}^\ast_v = 4\pi n \mathcal{C} \int^{\infty}_{0}dr~r^{2}/(1+i\zeta r^{6}); ~\zeta = \frac{|\Omega|^{2}}{4C_{6}\Gamma_{e}}.
\ee
We can write this integral as $\mathcal{C}^{\ast}_{v} = \mathcal{C}_{b}-i\mathcal{C}'_{b}$ and solved it to get, $|\mathcal{C}_{b}| = |\mathcal{C}'_{b}| = \frac{2}{3}(\mathcal{C}n\pi^{2}/\sqrt{2\zeta})$. Above, we have solved the scattering dynamics in the case where there was already a Rydberg excitation stored. In principle, we should also solve the dynamics without the first stored excitation. In this case, however, the excitations are completely independent of each other. We can then conveniently obtain the results for this situation by simple setting $\mathcal{V}_{kl} = 0$. Then from Eq. (\ref{5}) we get $\mathcal{C}^\ast_{v} = 0$ for a long photon pulse and hence $\mathcal{R}_{g} = 1$. 

To investigate the effect of pulses of a finite duration, we now consider the  bandwidth of the scattering coefficient. To do this, we perform a Taylor series expansion of the reflection coefficient about some central frequency $\omega_{0}$,
\bea
\label{7}
\mathcal{R}_{k}(\omega) & = & \mathcal{R}_{k}(\omega_{0})+\partial_{\omega}\mathcal{R}_{k}|_{\omega_{0}}(\omega-\omega_{0})+\frac{1}{2}\partial^{2}_{\omega}\mathcal{R}_{k}|_{\omega_{0}}(\omega-\omega_{0})^{2}.
\eea
Here we have kept upto the second order in the expansion. The above three terms in the expansion are described by,
\be
\label{8}
\mathcal{R}_{k}(\omega_{0}) = \left(\frac{2}{1+\mathcal{C}^{\ast}_{v}}-1\right) + 2i\frac{\omega_{0}}{\kappa}\frac{1}{(1+\mathcal{C}^{\ast}_{v})^{2}},
\ee
\bea
\label{9}
\partial_{\omega}\mathcal{R}_{k}|_{\omega_{0}} & = &\frac{-4\frac{\omega_{0}}{\kappa}\left(\frac{1}{\kappa}-\frac{\mathcal{C}^{\ast}_{\alpha_{v}}}{\Gamma_{e}}\right)}{\left(1+\mathcal{C}^{\ast}_{v}\right)^{3}} +\frac{2i\left(\frac{1}{\kappa}-\frac{\mathcal{C}^{\ast}_{\alpha_{v}}}{\Gamma_{e}}\right)}{\left(1+\mathcal{C}^{\ast}_{v}\right)^{2}},\\
\label{10}
\partial^{2}_{\omega}\mathcal{R}_{k}|_{\omega_{0}} & = &\frac{-4\left(\frac{1}{\kappa}-\frac{\mathcal{C}^{\ast}_{\alpha_{v}}}{\Gamma_{e}}\right)^{2}}{\left(1+\mathcal{C}^{\ast}_{v}\right)^{3}}+\frac{\frac{4\mathcal{C}^{\ast}_{v}}{\Gamma_{e}}\frac{\mathcal{C}^\ast_{\beta_{v}}}{\Gamma_{e}}\left[(\mathcal{C}^\ast_{v})^{2}-3\frac{\omega^{2}_{0}}{\kappa^{2}}\right]}{\left(1+\mathcal{C}^{\ast}_{v}\right)^{6}}+\frac{4\frac{\omega_{0}}{\kappa}\frac{\mathcal{C}^{\ast}_{\eta_{v}}}{\Gamma^{2}_{e}}}{\left(1+\mathcal{C}^{\ast}_{v}\right)^{4}}+\frac{4\frac{\mathcal{C}^{\ast}_{\chi_{v}}}{\Gamma^{2}_{e}}}{\left(1+\mathcal{C}^{\ast}_{v}\right)^{3}},\nonumber\\
&-4i&\left\{\frac{3\frac{\omega_{0}}{\kappa}\left(\frac{1}{\kappa}-\frac{\mathcal{C}^{\ast}_{\alpha_{v}}}{\Gamma_{e}}\right)^{2}}{\left(1+\mathcal{C}^{\ast}_{v}\right)^{4}}-\frac{\frac{\omega_{0}}{\kappa}\frac{\mathcal{C}^{\ast}_{\beta_{v}}}{\Gamma^{2}_{e}}\left[3(\mathcal{C}^{\ast}_{v})^{2}-\frac{\omega^{2}_{0}}{\kappa^{2}}\right]}{\left(1+\mathcal{C}^{\ast}_{v}\right)^{6}}+\frac{\frac{\mathcal{C}^{\ast}_{\eta_{v}}}{\Gamma^{2}_{e}}}{\left(1+\mathcal{C}^{\ast}_{v}\right)^{3}}+\frac{\frac{\omega_{0}}{\kappa}\frac{\mathcal{C}^{\ast}_{\chi_{v}}}{\Gamma^{2}_{e}}}{\left(1+\mathcal{C}^{\ast}_{v}\right)^{4}}\right\},\nonumber\\
\eea
with the parameters defined by, 
\bea
\label{part1a}
\mathcal{C}^\ast_{v} & = &\sum_{l}\frac{\mathcal{C}}{1-i\frac{\omega_{0}}{\Gamma_{e}}+\frac{|\Omega_{l}/2|^{2}}{i(\mathcal{V}_{kl}-\omega_{0})\Gamma_{e}}},\qquad \mathcal{C}^{\ast}_{\alpha_{v}} = \mathcal{C}\sum_{l}\frac{1+|\Omega_{l}/2|^{2}/(\mathcal{V}_{kl}-\omega_{0})^{2}}{\left[1-i\frac{\omega_{0}}{\Gamma_{e}}+\frac{|\Omega_{l}|^{2}}{i(\mathcal{V}_{kl}-\omega_{0})\Gamma_{e}}\right]^{2}}\\
\label{part1b}
\mathcal{C}^{\ast}_{\beta_{v}}& = &\mathcal{C}\sum_{l}\frac{1+|\Omega_{l}/2|^{2}/(\mathcal{V}_{kl}-\omega_{0})^{2}}{\left[1-i\frac{\omega_{0}}{\Gamma_{e}}+\frac{|\Omega_{l}/2|^{2}}{i(\mathcal{V}_{kl}-\omega_{0})\Gamma_{e}}\right]^{3}},\qquad
\mathcal{C}^{\ast}_{\eta_{v}} = \mathcal{C}\sum_{l}\frac{|\Omega_{l}/2|^{2}\Gamma_{e}/(\mathcal{V}_{kl}-\omega_{0})^{3}}{\left[1-i\frac{\omega_{0}}{\Gamma_{e}}+\frac{|\Omega_{l}/2|^{2}}{i(\mathcal{V}_{kl}-\omega_{0})\Gamma_{e}}\right]^{3}}\\
\label{part2}
\mathcal{C}^{\ast}_{\chi_{v}} & = &\mathcal{C} \sum_{l}\frac{|\Omega_{l}/2|^{2}\omega_{0}/(\mathcal{V}_{kl}-\omega_{0})^{3}}{\left[1-i\frac{\omega_{0}}{\Gamma_{e}}+\frac{|\Omega_{l}/2|^{2}}{i(\mathcal{V}_{kl}-\omega_{0})\Gamma_{e}}\right]^{3}}.
\eea
Assuming the central frequency of the incoming pulse to be on resonance, we set $\omega_0 = 0$ and hence Eqs. (\ref{8}-\ref{10}) become substantially simpler and are described by,  
\bea
\label{11}
\mathcal{R}_{k} &=& \left(\frac{2}{1+\mathcal{C}^{\ast}_{v}}-1\right),\\
\label{12}
\partial_{\omega}\mathcal{R}_{k}|_{\omega_{0} = 0} & = &\frac{2i\left(\frac{1}{\kappa}-\frac{\mathcal{C}^{\ast}_{\alpha_{v}}}{\Gamma_{e}}\right)}{\left(1+\mathcal{C}^{\ast}_{v}\right)^{2}},\\
\label{13}
\partial^{2}_{\omega}\mathcal{R}_{k}|_{\omega_{0}=0}& = &\frac{-4\left(\frac{1}{\kappa}-\frac{\mathcal{C}^{\ast}_{\alpha_{v}}}{\Gamma_{e}}\right)^{2}}{\left(1+\mathcal{C}^{\ast}_{v}\right)^{3}}+\frac{\frac{4\mathcal{C}^{\ast}_{v}}{\Gamma_{e}}\frac{\mathcal{C}^{\ast}_{\beta_{v}}}{\Gamma_{e}}(\mathcal{C}^\ast_{v})^{2}}{\left(1+\mathcal{C}^{\ast}_{v}\right)^{6}}+\frac{4\frac{\mathcal{C}^{\ast}_{\chi_{v}}}{\Gamma^{2}_{e}}}{\left(1+\mathcal{C}^{\ast}_{v}\right)^{3}}-4i\left\{\frac{\frac{\mathcal{C}^{\ast}_{\eta_{v}}}{\Gamma^{2}_{e}}}{\left(1+\mathcal{C}^{\ast}_{v}\right)^{3}}\right\}.
\eea
Note that for the case of no stored photon in the ensemble, we have $\mathcal{R}_{k} \rightarrow \mathcal{R}_{g}$ and one can also get the Taylor series expansion of $\mathcal{R}_{g}$ by setting $\mathcal{V}_{kl} = 0$. The expressions for such an expansion at resonance is the same as given by Eqs. (\ref{11} - \ref{13}) but now with the set of parameters, $\mathcal{C}^\ast_{v}\rightarrow \mathcal{C}^\ast, \mathcal{C}^{\ast}_{\alpha_{v}} \rightarrow \mathcal{C}^{\ast}_{\alpha}, \mathcal{C}^{\ast}_{\beta_{v}} \rightarrow \mathcal{C}^{\ast}_{\beta}, \mathcal{C}^{\ast}_{\eta_{v}}\rightarrow \mathcal{C}^{\ast}_{\eta}, \mathcal{C}^{\ast}_{\chi_{v} }\rightarrow \mathcal{C}^{\ast}_{\chi}$, where the new parameters correspond to Eqs. (\ref{part1a} -\ref{part2}) with $\mathcal{V}_{kl} = 0$. From the set of Eqs. (\ref{11}-\ref{13}), we see that the leading order dispersive contributions are scaled down by a factor of $(\mathcal{C}^\ast_{v})^{2}$ and $(\mathcal{C}^\ast_{v})^{3}$ when we compare the situation with and without stored excitation in the first pulse. Hence the spectrally narrowest feature is the width of the EIT resonance without a stored excitation, and this will thus be the limiting factor for the bandwidth.

We next analyze the behaviour of the parameters listed in Eqs. (\ref{part1a}-\ref{part2}) in different limits of operation. We can find the blockaded part by considering the limits $\mathcal{V}_{kl} \gg |\Omega_{l}/2|^{2}/\Gamma_{e}$, while the contribution from the remaining EIT medium is found in the limit $\mathcal{V}_{kl} \ll |\Omega_{l}/2|^{2}/\Gamma_{e}$. To get a feeling for the expression in Eqs. (\ref{part1a}-\ref{part2}), we separate them into contributions coming from the blockaded atoms and that from the rest of EIT medium,
\bea
\label{14}
\mathcal{C}^\ast_{v} & \approx &\mathcal{C}_{b}+i\mathcal{C}'_{b} = \sum_{l}\frac{\mathcal{C}}{\left[1+\frac{|\Omega/2|^{4}}{\mathcal{V}^{2}_{kl}\Gamma^{2}_{e}}\right]}+i\sum_{l}\frac{\mathcal{C}\frac{|\Omega/2|^{2}}{\mathcal{V}^{2}_{kl}\Gamma_{e}}}{\left[1+\frac{|\Omega/2|^{4}}{\mathcal{V}_{kl}\Gamma^{2}_{e}}\right]};\\
\label{15}
\mathcal{C}^{\ast}_{\alpha_{v}} & = &\mathcal{C}^{\ast\alpha}_{b}-N^{\alpha}_{EIT}\mathcal{C}\frac{\Gamma^{2}_{e}}{|\Omega/2|^{2}},\qquad\mathcal{C}^{\ast}_{\beta_{v}} = \mathcal{C}^{\ast\beta}_{b};\\
\label{16}
\mathcal{C}^{\ast}_{\eta_{v}} & = &\mathcal{C}^{\ast\eta}_{b}-iN^{\eta}_{EIT}\mathcal{C}\frac{\Gamma^{4}_{e}}{|\Omega/2|^{4}},\qquad\mathcal{C}^{\ast}_{\chi_{v}}= 0, 
\eea
where we have assumed all the Rabi frequencies to be equal such that $\Omega_{l} = \Omega$ and $\mathcal{C}^{\ast\alpha}_{v} = \sum_{l}\mathcal{C}/(1+|\Omega/2|^{2}/i\mathcal{V}_{kl}\Gamma_{e})^{2}, \mathcal{C}^{\ast\beta}_{v} = \mathcal{C}^{\ast\eta}_{v} = \sum_{l}\mathcal{C}/(1+|\Omega/2|^{2}/i\mathcal{V}_{kl}\Gamma_{e})^{3}$, which scale as the number of blocked atoms $\mathcal{C}_{b}$ while $N^{\alpha}_{EIT}, N^{\beta}_{EIT}$ scale as the number of remaining unblocked atoms $\sim N$. Similarly, for the case of no stored excitation, we get, 
\bea
\mathcal{C}^\ast & = &0,\qquad\mathcal{C}^{\ast}_{\alpha} = -N\mathcal{C}\frac{\Gamma^{2}_{e}}{|\Omega/2|^{2}},\qquad\mathcal{C}^{\ast}_{\beta} = 0;\\
\mathcal{C}^{\ast}_{\eta}& = &-iN\mathcal{C}\frac{\Gamma^{4}_{e}}{|\Omega/2|^{4}},\qquad\mathcal{C}^\ast_{\chi} = 0.
\eea
Note that since in this case there is no Rydberg excitation blockade, only the EIT medium contributes and all the terms arising due to blockade are zero.

\section{Choi-Jamiolkowski fidelity}
The Choi-Jamiolkowski (CJ) fidelity is a measure of how close two given quantum 
mechanical processes are. The idea is to apply each process to a particular entangled 
state and then calculate the fidelity between the two output states. Specifically, 
we assume that the two processes are described by the superoperators 
$\mathcal{U}$ and $\mathcal{V}$. The superoperator $\mathcal{U}$ represents 
the ideal process that we want to accomplish and is assumed to be unitary. 
Hence, its action on some density matrix $\rho$ can be be written as
\begin{gather}\label{superoperator_U_Kraus}
\mathcal{U}(\rho)=U\rho U^\dagger,
\end{gather}
where $U$ is a unitary operator. The actual physical implementation is 
represented by the completely positive trace preserving superoperator 
$\mathcal{V}$. In general, it admits a Kraus (operator-sum) decomposition
\begin{gather}\label{superoperator_V_Kraus}
\mathcal{V}(\rho)=\sum_l V_l\rho V_l^\dagger
\end{gather}
with $\sum_l V_l^\dagger V_l=I$ ($I$ is the identity operator). If we 
separate out the ``no jump'' evolution with the effective non-Hermitian 
Hamiltonian $\mathcal{H}$ in Eq. \eqref{superoperator_V_Kraus}, we can write
\begin{gather}\label{superoperator_V_Kraus_eff_Hamiltonian}
\mathcal{V}(\rho)= V\rho V^\dagger
+\sum_l K_l\rho K_l^\dagger,
\end{gather}
where $V=\exp(-i\mathcal{H}t_\text{f}/\hbar)$ with $t_\text{f}$ being the time 
it takes to accomplish the wanted operation. The operators $K_l$ form the 
Kraus decomposition of the part of the evolution where at least one quantum 
jump occurs.

To find the CJ fidelity, we consider the superoperators 
$\mathcal{I}\otimes\mathcal{U}$ and $\mathcal{I}\otimes\mathcal{V}$ that are 
tensor products of the original ones with the identity superoperator 
$\mathcal{I}$. We pick an orthonormal basis set $\{|j\rangle\}$ for the $d$-dimensional 
Hilbert space that $\mathcal{U}$ and $\mathcal{V}$ act on. Now we can define the state
$|\Phi\rangle=\sum_j|j\rangle|j\rangle/\sqrt{d}$ that is an element of the 
original Hilbert space tensored with a copy of itself. Note that $|\Phi\rangle$ 
is a maximally entangled state of these two copies. We will only consider a two-qubit gate, so that $d=4$ in the above.

After applying $\mathcal{I}\otimes\mathcal{U}$ and 
$\mathcal{I}\otimes\mathcal{V}$ onto the density matrix 
$|\Phi\rangle\langle\Phi|$, we get a pair of new states
\begin{eqnarray}
&&\rho_\mathcal{U}=[\mathcal{I}\otimes\mathcal{U}]|\Phi\rangle\langle\Phi | =(I\otimes U)|\Phi\rangle\langle\Phi |(I\otimes U^\dagger),\\
&&\rho_\mathcal{V}=[\mathcal{I}\otimes\mathcal{V}]|\Phi\rangle\langle\Phi | =(I\otimes V)|\Phi\rangle\langle\Phi |(I\otimes V^\dagger)
+\sum_{l}(I\otimes K_l)|\Phi\rangle\langle\Phi |(I\otimes K_l^\dagger).
\end{eqnarray}
The CJ fidelity is defined to be the fidelity of these two states. Since $\rho_{\mathcal{U}}$ is a pure state, we get
\begin{gather}\label{CJ_fidelity_general}
\begin{aligned}
F_\text{CJ}=F(\rho_\mathcal{U},\rho_\mathcal{V})
&=\langle\Phi|(I\otimes U^\dagger)\rho_\mathcal{V}(I\otimes U)|\Phi\rangle\\
&=|\langle\Phi|(I\otimes U^\dagger V)|\Phi\rangle|^2
+\sum_{l}|\langle\Phi|(I\otimes U^\dagger K_l)|\Phi\rangle|^2.
\end{aligned}
\end{gather}
\section{Storage and retrieval}
The full physical process to implement the controlled-phase gate consists of 
storage of one photon, scattering of the second one, and retrieval of the 
first. The theory of storage and retrieval with an ensemble in a cavity is 
well established. In suitable regimes these results show that we have a 
mapping between a single mode of the atomic ensemble and a specific incoming 
or outgoing optical mode, and all other modes will be uncoupled 
\cite{Gorshkov07}. Hence, the process of storage is described by a 
single parameter, which is the storage efficiency of a single incoming photon 
to create a specific spin wave
\begin{gather}\label{S_definition}
|S\rangle = \sum_k \alpha_{k}|g^{N-1}, r'_k\rangle.
\end{gather}
After scattering of the 
second photon, this spin wave will get multiplied by the reflection coefficient 
of the second photon such that it becomes 
\begin{gather*}
|S_\mathcal{R}\rangle=\sum_k \alpha_{k}\mathcal{R}_k(\omega)|g^{N-1}, r'_k\rangle,
\end{gather*}
where $\omega$ is the frequency of the second photon. Note here that the 
scattering coefficient may depend on which atom the first photon was stored 
in since different atoms may experience different degrees of blockade. For 
the retrieval, the cavity maps the particular spin wave 
\eqref{S_definition} to a specific temporal mode. Hence, the amplitude of 
the retrieved photon is given by the shape of that temporal mode multiplied by the overlap
\begin{gather*}
\langle S|S_\mathcal{R}\rangle
=\sum_k |\alpha_{k}|^2 \mathcal{R}_k(\omega).
\end{gather*}
In general, the retrieved wavepacket will also need to multiplied by the square 
root of the overall storage and retrieval efficiency, but we neglect this in our 
analysis.

For the fidelity calculations, one would need to calculate the overlap of the 
retrieved photon wavepackets corresponding to $|S\rangle$ and 
$|S_\mathcal{R}\rangle$. However, by the discussion above, the overlap of the 
photon wavepackets will be equal to the overlap of the spin waves $|S\rangle$ 
and $|S_\mathcal{R}\rangle$. Hence, in the calculations below, we will directly 
calculate the fidelities by projecting the spin waves instead of analysing the retrieval.

\section{Fidelity in the single-rail encoding}
In the single-rail encoding the computational basis is
\begin{gather}\label{single_rail_initial_states}
\begin{aligned}
&|00(t)\rangle = |g^N\rangle|\text{\O}\rangle,\\
&|01(t)\rangle = |g^N\rangle
\int d\omega \phi(\omega) \hat{a}^{\dagger}_{\omega}e^{-i\omega t}|\text{\O}\rangle,\\
&|10(t)\rangle = \sum_k \alpha_{k}|g^{N-1}, r'_k\rangle|\text{\O}\rangle,\\
&|11(t)\rangle = \sum_k \alpha_{k}|g^{N-1}, r'_k\rangle
\int d\omega \phi(\omega) \hat{a}^{\dagger}_{\omega}e^{-i\omega t}|\text{\O}\rangle.\\
\end{aligned}
\end{gather}
Note that this basis is time dependent due to the free evolution phase 
$\exp(-i\omega t)$. Hence, we define the ideal operation $\mathcal{U}$ 
such that it includes this free evolution phase. Specifically, if we denote 
the computational basis states at the initial time $t=0$ by omitting the time 
variable, i.e.  $|jj'\rangle=|jj'(t=0)\rangle$ ($j,j'\epsilon {0,1}$), then the 
ideal operation of the controlled-phase gate is given by
\begin{gather}\label{perfect_controlled_phase_gate}
U|00\rangle = |00(t_\text{f})\rangle,\quad
U|01\rangle = |01(t_\text{f})\rangle,\quad
U|10\rangle = |10(t_\text{f})\rangle,\quad
U|11\rangle = -|11(t_\text{f})\rangle.
\end{gather}

Using the computational basis \eqref{single_rail_initial_states} we can write
\begin{gather*}
|\Phi\rangle = \frac{1}{2}\del{
|00\rangle|00\rangle
+|01\rangle|01\rangle
+|10\rangle|10\rangle
+|11\rangle|11\rangle
}.
\end{gather*}
Inserting this specific form of $|\Phi\rangle$ into 
\eqref{CJ_fidelity_general} we obtain
\begin{gather}\label{CJ_fidelity_computational_basis}
\begin{aligned}
F_\text{CJ}=
&\frac{1}{16} \envert{
\langle 00|U^\dagger V|00\rangle
+\langle 01|U^\dagger V|01\rangle
+\langle 10|U^\dagger V|10\rangle
+\langle 11|U^\dagger V|11\rangle
}^2\\
&+\frac{1}{16}\sum_{l} \envert{
\langle 00|U^\dagger K_l|00\rangle
+\langle 01|U^\dagger K_l|01\rangle
+\langle 10|U^\dagger K_l|10\rangle
+\langle 11|U^\dagger K_l|11\rangle
}^2.
\end{aligned}
\end{gather}

For the operators $K_l$, we assume that 
$\langle jj'(t_\text{f})|K_l|jj'\rangle = 0$, where $j,j'\in\cbr{0,1}$. 
Physically, this assumption means that if a quantum jump (incoherent decay) 
occurs, the given basis will switch to another state of the physical system 
(possibly even one of the other basis states) but can never be driven back to 
the original state. I.e. if a photon is lost, it will result in a vacuum 
output, and thus is does not give an overlap with the original state. Under 
this assumption, we only need to compute the dynamics due to the non-Hermitian 
Hamiltonian. The detailed calculation is presented in 
Sec.~\ref{sec_reflection_coefficient}. In essence, the result is that the 
dynamics of the operator $V$ can be described by the scattering relations
\begin{gather}\label{V_action_on_initial_states}
\begin{aligned}
&V|00\rangle = |g^N\rangle|\text{\O}\rangle,\\
&V|01\rangle = |g^N\rangle
\int d\omega \mathcal{R}_g(\omega) \phi(\omega) \hat{a}^{\dagger}_{\omega}e^{-i\omega t_\text{f}}
|\text{\O}\rangle,\\
&V|10\rangle = \sum_k \alpha_{k}|g^{N-1}, r'_k\rangle|\text{\O}\rangle,\\
&V|11\rangle = \int d\omega \sum_k \alpha_{k}\mathcal{R}_k(\omega)
\phi(\omega) \hat{a}^{\dagger}_{\omega}e^{-i\omega t_\text{f}}
|g^{N-1}, r'_k\rangle|\text{\O}\rangle.\\
\end{aligned}
\end{gather}
Gathering all the formulas in this section, the CJ fidelity becomes
\begin{gather}\label{CJ_fidelity_single_rail}
F_\text{CJ}
=\frac{1}{16}\envert{
2 + \int d\omega |\phi(\omega)|^2 \mathcal{R}_g(\omega) 
-\int d\omega|\phi(\omega)|^2 \sum_k |\alpha_{k}|^2 \mathcal{R}_k(\omega)}^2.
\end{gather}

\section{Fidelity in the dual-rail encoding}

\begin{figure}[hbt]
\begin{center}
 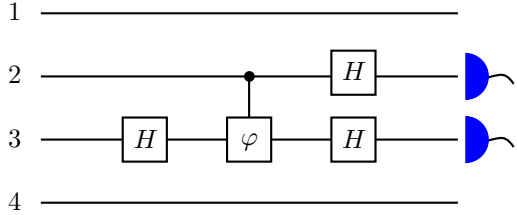
\end{center}
\caption{(a) The circuit diagram of the entanglement swap operation. The 
numbers at the left edge indicate the label of the subsystem (qubit). In the 
circuit, the Hadamard gates are denoted by $H$ and the controlled-phase gate 
is denoted
by $\varphi$.}
\label{fig_entanglement_swap_circuit}
\end{figure}

In this section we calculate both the CJ fidelity and the entanglement 
swap fidelity for the dual-rail encoding and show how they relate to each 
other. The circuit diagram of the entanglement swap operation is shown in 
Fig.~\ref{fig_entanglement_swap_circuit}.

The entanglement swap operation consists of evolution of the initial state 
(which is unitary in the ideal case) and a subsequent measurement. The 
evolution can be decomposed into a controlled-phase gate and Hadamard gates. 
If the Hadamard gates are assumed to be ideal, then the CJ fidelity of the 
whole evolution is equal to the CJ fidelity of the controlled-phase gate. We 
are going to use this fact in relating the CJ fidelity to the entanglement 
swap fidelity.

The abstract definition of the CJ fidelity does not make any reference to a 
particular basis. In this section, in addition to the computational basis, we 
will also use the Bell basis, since it is the natural choice for the 
entanglement swap operation. The Bell states are
\begin{gather}\label{Bell_states_definition}
\begin{aligned}
&|\phi^{00}\rangle=|\phi^+\rangle=\frac{1}{\sqrt{2}}\del{|00\rangle+|11\rangle},\\
&|\phi^{01}\rangle=|\psi^+\rangle=\frac{1}{\sqrt{2}}\del{|01\rangle+|10\rangle},\\
&|\phi^{10}\rangle=|\phi^-\rangle=\frac{1}{\sqrt{2}}\del{|00\rangle-|11\rangle},\\
&|\phi^{11}\rangle=|\psi^-\rangle=\frac{1}{\sqrt{2}}\del{|01\rangle-|10\rangle}.
\end{aligned}
\end{gather}
In addition to the conventional names, we also give numbers to the Bell 
states, which will allow us to express summations in a simple way below.

For the entanglement swap circuit of Fig.~\ref{fig_entanglement_swap_circuit}, the 
initial state is one Bell pair $|\phi^+\rangle_{12}$ between subsystems 1 
and 2 and another Bell pair $|\phi^+\rangle_{34}$ between subsystems 3 and 
4. Note that this initial state can be written as
\begin{gather*}
|\phi^+\rangle_{12}|\phi^+\rangle_{34}
=\frac{1}{2}\sum_{j,j'=0}^{1}
|\phi^{jj'}\rangle_{14}|\phi^{jj'}\rangle_{23}
=|\Phi\rangle_{1423}.
\end{gather*}
This is exactly the state that is used as the input for the calculation of the 
CJ fidelity expressed in the Bell basis. After evolution of subsystems 2 
and 3 as shown by the circuit and a measurement (the two detectors to the 
right), a Bell pair between subsystems 1 and 4 is established.

The practical implementation of the above circuit is shown in Fig.~1(c) of the 
main text. Whereas Fig.~\ref{fig_entanglement_swap_circuit} displays the 
extended four-qubit Hilbert space required for the calculation of the CJ and 
entanglement swap fidelity, Fig.~1(c) only displays the two central subsystems 
(2 and 3), but each of the two subsystems are represented by the photon being 
in two distinct modes.

We define $\hat{a}_{0,\omega}^\dagger$ to be the creation operator for subsystem 
3 in state $|0\rangle$ with frequency $\omega$, and 
$\hat{a}_{1,\omega}^\dagger$ to be 
the creation operator for state $|1\rangle$. For notational convenience, we define 
the states $|0_\omega\rangle_3=\hat{a}_{0,\omega}^\dagger|\text{\O}\rangle$ and 
$|1_\omega\rangle_3=\hat{a}_{1,\omega}^\dagger|\text{\O}\rangle$. Then in the dual-rail encoding, 
the computational basis is
\begin{gather}\label{dual_rail_initial_states}
\begin{aligned}
&|0\rangle_2
=\sum_k \alpha_k |g^{N-1}, r'_k\rangle_0,\\
&|1\rangle_2
=\sum_k \alpha_k |g^{N-1}, r'_k\rangle_1,\\
&|0(t)\rangle_3
=\int d\omega \phi(\omega)e^{-i\omega t}|0_\omega\rangle_3,\\
&|1(t)\rangle_3
=\int d\omega \phi(\omega)e^{-i\omega t}|1_\omega\rangle_3.
\end{aligned}
\end{gather}
Here, $|g^{N-1}, r'_k\rangle_0$ are the states of the memory (the ensemble 
which does not interact with the second photon), and $|g^{N-1}, r'_k\rangle_1$ 
are the states of the cavity from which the second photon is scattered (see 
Fig.~1(c) of the main text). These two states correspond to 
subsystem 2 of Fig.~\ref{fig_entanglement_swap_circuit}. Subsystem 3 is 
encoded in photonic states which are not stored but only scattered. Note that 
all of the resulting computational basis states $|00\rangle_{23}$, 
$|01\rangle_{23}$, $|10\rangle_{23}$ and $|11\rangle_{23}$ physically 
correspond to having two excitations. Hence, it allows for simple means of 
error detection: if less than two excitations are present at the end of the 
evolution, we know that an error has occured.  In the dual-rail basis, the 
action of the operator $V_{23}$ that corresponds to the physical 
implementation of the controlled-phase gate can be written
\begin{gather}\label{dual_rail_V_action_on_initial_states}
\begin{aligned}
&V_{23}|00\rangle_{23}
=\sum_k \alpha_k |g^{N-1}, r'_k\rangle_0
\int d\omega \phi(\omega)e^{-i\omega t_\text{f}}|0_\omega\rangle_3,\\
&V_{23}|01\rangle_{23}
=\sum_k \alpha_k |g^{N-1}, r'_k\rangle_0
\int d\omega \mathcal{R}_g(\omega)\phi(\omega)e^{-i\omega t_\text{f}}
|1_\omega\rangle_3,\\
&V_{23}|10\rangle_{23}
=\sum_k \alpha_k |g^{N-1}, r'_k\rangle_1
\int d\omega \phi(\omega)e^{-i\omega t_\text{f}}|0_\omega\rangle_3,\\
&V_{23}|11\rangle_{23}
=\int d\omega \sum_k \alpha_k \mathcal{R}_k(\omega)
\phi(\omega)e^{-i\omega t_\text{f}}
|g^{N-1}, r'_k\rangle_1|1_\omega\rangle_3.
\end{aligned}
\end{gather}

For the operators corresponding to the full evolution of the circuit of 
Fig.~\ref{fig_entanglement_swap_circuit}, we also need to describe the 
Hadamard operators. In the dual-rail encoding, the Hadamard operations are 
obtained by impinging the photons on beamsplitters which work on all frequency 
components separately. This is important for subsystem~3 (the 
scattered photon), since the 
frequency components will be multiplied with, in general, different reflection 
coefficients $\mathcal{R}_g(\omega)$ and $\mathcal{R}_k(\omega)$ depending on 
the input state. Hence the definition of the Hadamard operator here needs to 
be per frequency component, i.e. $H_3|0_\omega\rangle_3=(|0_\omega\rangle_3+|1_\omega\rangle_3)/\sqrt{2}$ and
$H_3|1_\omega\rangle_3=(|0_\omega\rangle_3-|1_\omega\rangle_3)/\sqrt{2}$. For 
subsystem 2, the single mode retrieval precludes any such difference in 
the mode shape for photons that are incident on the beamsplitters. Hence we can define the Hadamard operators to 
act on the spin wave states directly, 
$H_2|0\rangle_2=(|0\rangle_2+|1\rangle_2)/\sqrt{2}$ and
$H_2|1\rangle_2=(|0\rangle_2-|1\rangle_2)/\sqrt{2}$.

In analogy with Eqs. \eqref{superoperator_U_Kraus} and 
\eqref{superoperator_V_Kraus_eff_Hamiltonian}, we define the superoperators 
$\tilde{\mathcal{U}}$ and $\tilde{\mathcal{V}}$ for the ideal and the real 
version of the circuit of Fig.~\ref{fig_entanglement_swap_circuit}. They can 
be written as
\begin{gather*}
\tilde{\mathcal{U}}_{23}(\rho)=\tilde{U}_{23}\rho \tilde{U}_{23}^\dagger,
\end{gather*}
and
\begin{gather*}
\tilde{\mathcal{V}}_{23}(\rho)= \tilde{V}_{23}\rho \tilde{V}_{23}^\dagger
+\sum_l \tilde{K}_{l,23}\rho \tilde{K}_{l,23}^\dagger,
\end{gather*}
where
\begin{gather}\label{operators_with_tilde}
\begin{aligned}
&\tilde{U}_{23}=(H_2\otimes H_3)U_{23}(I_2\otimes H_3),\\
&\tilde{V}_{23}=(H_2\otimes H_3)V_{23}(I_2\otimes H_3),\\
&\tilde{K}_{l,23}=(H_2\otimes H_3)K_{l,23}(I_2\otimes H_3).
\end{aligned}
\end{gather}
Note that with the definitions \eqref{perfect_controlled_phase_gate}, 
\eqref{Bell_states_definition} and \eqref{operators_with_tilde}, it holds that 
$\tilde{U}_{23}|\phi^{jj'}\rangle = |jj'(t_\text{f})\rangle$.

In this setting, not only the input states used for the entanglement swapping match the ones used for the CJ 
fidelity, also the actual operation itself has the same form: an identity 
operation acting on 
subsystems 1 and 4, while subsystems 2 and 3 are evolved according to 
either $\tilde{\mathcal{U}}$ or $\tilde{\mathcal{V}}$. The two output states are then
\begin{align*}
&\rho_{\tilde{\mathcal{U}}}
=[\mathcal{I}_{14}\otimes\tilde{\mathcal{U}}_{23}]
\del{|\Phi\rangle\langle\Phi|}
=(I_{14}\otimes \tilde{U}_{23})|\Phi\rangle\langle\Phi|
(I_{14}\otimes \tilde{U}_{23}^\dagger),\\
&\rho_{\tilde{\mathcal{V}}}
=[\mathcal{I}_{14}\otimes\tilde{\mathcal{V}}_{23}]
\del{|\Phi\rangle\langle\Phi|}
=(I_{14}\otimes \tilde{V}_{23})|\Phi\rangle\langle\Phi|
(I_{14}\otimes \tilde{V}_{23}^\dagger)
+\sum_{l}(I_{14}\otimes \tilde{K}_{l,23})|\Phi\rangle\langle\Phi|
(I_{14}\otimes \tilde{K}_{l,23}^\dagger).
\end{align*}

Now we want to use the error detection property of the dual-rail 
encoding. We define the projection operators
\begin{gather}\label{projection_operator_jj}
\hat{P}_{jj'}=I_{14}\otimes|j\rangle\langle j|_2\otimes 
\del{\int d\omega |j'_\omega\rangle\langle j'_\omega|_3}
\end{gather}
where $j,j'\in\cbr{0,1}$, and we also define their sum
\begin{gather}\label{projection_operator_sum}
\hat{P}=\sum_{j,j'=0}^{1} \hat{P}_{jj'}.
\end{gather}
The projection operators of Eq.~\eqref{projection_operator_jj} correspond to measuring the states $|jj'\rangle_{23}$ on the detectors of circuit of Fig.~\ref{fig_entanglement_swap_circuit}. Note that for subsystem 3, we 
project onto the entire subspace that is spanned by the states 
$|j'_\omega\rangle$ instead of choosing a particular mode. This is equivalent 
to the assumption that all frequency components contribute to the probability of 
a ``click'' on the detector. On the other hand, the operator of 
Eq.~\eqref{projection_operator_sum} has a less clear physical interpretation.  
Formally, it projects a given state onto the subspace with two excitations. 
The motivation for defining such an operator is to be able to relate the CJ 
fidelity to the entanglement swap fidelity as we will see below. Since the 
entanglement swap fidelity can only be 
understood as a conditional fidelity (conditioned on the measurement outcomes 
corresponding to the operators of Eq.~\eqref{projection_operator_jj}), the 
CJ fidelity also needs to be conditional.

Let us begin with the calculation of the entanglement swap fidelity. Using the 
states after the measurement has taken place
\begin{gather*}
\rho_{jj'}
=\frac{\hat{P}_{jj'}\rho_{\tilde{\mathcal{V}}}\hat{P}_{jj'}^\dagger}
{\tr(\hat{P}_{jj'}\rho_{\tilde{\mathcal{V}}}\hat{P}_{jj'}^\dagger)},
\end{gather*}
we can define the conditional fidelities for the entanglement swap
\begin{gather}\label{entanglement_swap_fidelity_jj}
F_{jj'}=\langle\phi^{jj'}|\tr_{23}(\rho_{jj'})|\phi^{jj'}\rangle_{14}.
\end{gather}
Here, we take the trace over subsystems 2 and 3, since the relevant 
question is how close subsystems 1 and 4 are to a particular Bell 
pair. The trace can be written as
\begin{gather*}
\tr_{23}(\rho_{jj'})
=\sum_{n,n'=0}^{1} \int d\omega
\langle n|_2\langle n'_\omega|_3\rho_{jj'}|n'_\omega\rangle_3|n\rangle_2
=\frac{1}{\tr(\hat{P}_{jj'}\rho_{\tilde{\mathcal{V}}}\hat{P}_{jj'}^\dagger)}
\int d\omega
\langle j|_2\langle j'_\omega|_3\rho_{\tilde{\mathcal{V}}}
|j'_\omega\rangle_3|j\rangle_2.
\end{gather*}
Projecting $\rho_{\tilde{\mathcal{V}}}$ onto $|\phi^{jj'}\rangle_{14}$ we 
obtain
\begin{gather*}
\langle \phi^{jj'}|\rho_{\tilde{\mathcal{V}}}
|\phi^{jj'}\rangle_{14}
=\frac{1}{4}\tilde{V}_{23}|\phi^{jj'}\rangle_{23}\langle\phi^{jj'}|_{23}
\tilde{V}_{23}^\dagger
+\frac{1}{4}\sum_{l}\tilde{K}_{l,23}|\phi^{jj'}\rangle_{23}\langle\phi^{jj'}|_{23}
\tilde{K}_{l,23}^\dagger.
\end{gather*}
For the dual-rail encoding, we have a stronger assumption about the operators 
$K_l$ than for the single-rail encoding. We are going to assume that 
$\langle n|_2 \langle n'_\omega|_3K_l|jj'\rangle = 0$, where 
$n,n',j,j'\in\cbr{0,1}$. Physically, this assumption means that the decay 
processes take the state out of the computational basis entirely, since any 
such decay will reduce the number of the total excitations to less than two. 
Then the expression for the fidelity \eqref{entanglement_swap_fidelity_jj} 
becomes
\begin{gather*}
F_{jj'}
=\frac{1}{4\tr(\hat{P}_{jj'}\rho_{\tilde{\mathcal{V}}}\hat{P}_{jj'}^\dagger)}
\int d\omega
\envert{\langle j|_2\langle j'_\omega|_3(H_2\otimes H_3)V_{23}(I_2\otimes H_3)|\phi^{jj'}\rangle_{23}}^2.
\end{gather*}
with the trace in the denominator given by
\begin{gather*}
\tr(\hat{P}_{jj'}\rho_{\tilde{\mathcal{V}}}\hat{P}_{jj'}^\dagger)
=\frac{1}{4}\sum_{n,n'=0}^{1}\int d\omega
\envert{\langle n|_2\langle n'_\omega|_3(H_2\otimes H_3)V_{23}(I_2\otimes H_3)|\phi^{jj'}\rangle_{23}}^2.
\end{gather*}
For all $j$ and $j'$ we get
\begin{gather*}
\tr(\hat{P}_{jj'}\rho_{\tilde{\mathcal{V}}}\hat{P}_{jj'}^\dagger)
=\frac{1}{16} \int d\omega |\phi(\omega)|^2
\del{2+|\mathcal{R}_g(\omega)|^2+\envert{\sum_k |\alpha_{k}|^2 \mathcal{R}_k(\omega)}^2}
\end{gather*}
and
\begin{gather}\label{entanglement_swap_fidelity_jj_final}
F_\text{swap}=F_{jj'}=\frac{1}{16P_\text{suc}}\int d\omega|\phi(\omega)|^2
\envert{2+\mathcal{R}_g(\omega)-\sum_k |\alpha_{k}|^2 \mathcal{R}_k(\omega)}^2,
\end{gather}
where we have defined the success probability $P_\text{suc}
=\sum_{j,j'=0}^{1}\tr(\hat{P}_{jj'}\rho_{\tilde{\mathcal{V}}}
\hat{P}_{jj'}^\dagger)$, i.e.
\begin{gather}\label{P_suc_definition}
P_\text{suc}=\frac{1}{4} \int d\omega |\phi(\omega)|^2
\del{2+|\mathcal{R}_g(\omega)|^2+\envert{\sum_k |\alpha_{k}|^2 \mathcal{R}_k(\omega)}^2}.
\end{gather}

Now we look at the conditional CJ fidelity. Using the state
\begin{gather}\label{conditioned_rho_tilde_V}
\rho_{\tilde{\mathcal{V}}}'
=\frac{\hat{P}\rho_{\tilde{\mathcal{V}}}\hat{P}^\dagger}
{\tr(\hat{P}\rho_{\tilde{\mathcal{V}}}\hat{P}^\dagger)}
\end{gather}
we can define the conditional CJ fidelity as 
$F'_\text{CJ}=F(\rho_{\tilde{\mathcal{U}}},
\rho_{\tilde{\mathcal{V}}}')$.
By the cyclicity and linearity of the trace, we have
\begin{gather*}
\tr(\hat{P}\rho_{\tilde{\mathcal{V}}}\hat{P}^\dagger)
=\sum_{j,j'=0}^{1}\tr(\hat{P}_{jj'}\rho_{\tilde{\mathcal{V}}}
\hat{P}_{jj'}^\dagger)=P_\text{suc}.
\end{gather*}
The projection operator $\hat{P}$ has no effect on the states $|\Phi\rangle$, 
hence $F(\rho_{\tilde{\mathcal{U}}},
\hat{P}\rho_{\tilde{\mathcal{V}}}\hat{P}^\dagger)
=F(\rho_{\tilde{\mathcal{U}}},
\rho_{\tilde{\mathcal{V}}})$, and the analysis reduces to finding the
unconditional CJ fidelity and dividing by the success probability $P_\text{suc}$. The final result is
\begin{gather}\label{CJ_fidelity_dual_rail}
F'_\text{CJ}=\frac{1}{16P_\text{suc}}\envert{
2 + \int d\omega |\phi(\omega)|^2 \mathcal{R}_g(\omega) 
-\int d\omega|\phi(\omega)|^2 \sum_k |\alpha_{k}|^2 \mathcal{R}_k(\omega)}^2.
\end{gather}

Comparing Eqs. \eqref{entanglement_swap_fidelity_jj_final} and 
\eqref{CJ_fidelity_dual_rail} we see that the only difference is the order of 
integration and taking the absolute value. Thus in general, we have 
$F'_\text{CJ}\leq F_\text{swap}$. If the bandwidth of the second photon is narrow 
compared to the the frequency variations of $\mathcal{R}_g$ and $\mathcal{R}_k$, 
then the two fidelity measures become equal. The reason for this similarity is 
that the two measures consider the same input, but they do not consider 
exactly the same output. For the CJ fidelity the question we ask is what is 
the output with a particular mode, which we for simplicity take to be the same 
as the input mode. Possibly the CJ fidelity can therefore be increased by 
considering a more appropriate output mode. For the swapping fidelity we on 
the other hand consider everything which is incident on the photodetectors 
regardless of the temporal mode and hence this fidelity is higher.

\section{The gate fidelities for the Rydberg controlled-phase gate}
In the single-rail case, we evaluate the CJ fidelity \eqref{CJ_fidelity_single_rail}. To evaluate the quantity inside the modulus square we expand it and use Eq. (\ref{7}) and the corresponding expansion for $\mathcal{R}_{g}$ to get,
\be
\label{18}
\Delta\mathcal{R} = \mathcal{R}_{g} - \sum_{k}|\alpha_{k}|^{2}\mathcal{R}_{k} = \frac{2\mathcal{C}_{b}+2i\mathcal{C}'_{b}}{(1+\mathcal{C}_{b})+i\mathcal{C}'_{b}}
\ee
\be
\label{19}
\Delta\mathcal{R}' =\left[\mathcal{R}'_{g} - \sum_{k}|\alpha_{k}|^{2}\mathcal{R}'_{k}\right] = 2i\left(\frac{1}{\kappa}+\frac{N\mathcal{C}\Gamma_{e}}{|\Omega/2|^{2}}\right)-2i\left(\frac{1}{\kappa}+\frac{N^{\alpha}_{EIT}\mathcal{C}\Gamma_{e}}{|\Omega/2|^{2}}\right)\frac{1}{(1+\mathcal{C}_{b})^2}+\frac{\frac{2i\mathcal{C}^{\ast\alpha}_{b}}{\Gamma_{e}}}{(1+\mathcal{C}_{b})^2}
\ee
\bea
\label{20}
\Delta\mathcal{R}'' & = &\left[\mathcal{R}''_{g} - \sum_{k}|\alpha_{k}|^{2}\mathcal{R}''_{k}\right] = -4\left(\frac{1}{\kappa}+\frac{N\mathcal{C}\Gamma_{e}}{|\Omega/2|^{2}} \right)^{2}+4\left(\frac{1}{\kappa}+\frac{N^{\alpha}_{EIT}\mathcal{C}\Gamma_{e}}{|\Omega/2|^{2}} \right)^{2}\frac{1}{(1+\mathcal{C}_{b})^{3}}\nonumber\\
&+&4\frac{(\frac{\mathcal{C}^{\ast\alpha}_{b}}{\Gamma_{e}})^2}{(1+\mathcal{C}_{b})^{3}}-4\frac{N\Gamma^{2}_{e}\mathcal{C}}{|\Omega/2|^{4}}+4\frac{N^\eta_{EIT}\Gamma^{2}_{e}\mathcal{C}}{|\Omega/2|^{4}}\frac{1}{(1+\mathcal{C}_{b})^{3}}-4\frac{\frac{\mathcal{C}^{\ast\beta}_{b}}{\Gamma^{2}_{e}}}{\mathcal{C}_{b}^{3}}+4i\frac{\mathcal{C}^{\ast\eta}_{b}}{\Gamma^{2}_{e}(1+\mathcal{C}_{b})^{3}}
\eea
In deriving the above expressions, we have assumed that the ensemble is homogeneous and that the potential is isotropic. Hence we have dropped the index $k$ from $\mathcal{V}_{kl}$. We can then do the sum over $k$ and given that $\alpha_{k}$ are normalized, we have $\sum_{k}|\alpha_{k}|^{2} = 1$ in the above expressions. 

A closer look at Eq. (\ref{CJ_fidelity_single_rail}) suggests a further simplification which gives us,
\bea
\label{21}
F_\text{CJ} = \frac{1}{16}\left(4+|\Delta\mathcal{R}|^{2}+4\mathbf{Re}[\Delta\mathcal{R}]
+2\mathbf{Re}[\Delta\mathcal{R}''](\Delta\omega)^{2}
+\mathbf{Re}[\Delta\mathcal{R}\Delta\mathcal{R}''^\ast](\Delta\omega)^{2}\right)
\eea
where we have assumed narrow bandwidth of the pulse and defined the variance of the incoming pulse as $(\Delta\omega)^{2} = \int d\omega |\phi(\omega)|^{2}(\omega-\omega_{0})^{2}$. Substituting Eqs. (\ref{18} - \ref{20}) in Eq. (\ref{21}) and assuming that $\mathcal{C}_{b},\mathcal{C}'_{b} \gg 1$ and $\mathcal{C}^{\ast\beta}_{b}, \mathcal{C}^{\ast\alpha}_{b} < \mathcal{C}^{2}_{b}$, we get,
\bea
\label{22}
F_\text{CJ} &\simeq&\left[1 -\frac{(1+\mathcal{C}_{b})}{(1+\mathcal{C}_{b})^{2}+\mathcal{C}^{'2}_{b}}\right]+\frac{1}{4[(1+\mathcal{C}_{b})^{2}+\mathcal{C}^{'2}_{b}]}-\frac{N\mathcal{C}\Gamma^{2}_{e}}{2|\Omega/2|^{4}}(\Delta\omega)^{2}\left(1+\frac{\mathcal{C}_{b}(1+\mathcal{C}_{b})+\mathcal{C}^{'2}_{b}}{(1+\mathcal{C}_{b})^{2}+\mathcal{C}^{'2}_{b}}\right)\nonumber\\
&-&\frac{1}{2}\left(\frac{1}{\kappa}+\frac{N\mathcal{C}\Gamma_{e}}{|\Omega/2|^{2}}\right)^{2}\left(1+\frac{\mathcal{C}_{b}(1+\mathcal{C}_{b})+\mathcal{C}^{'2}_{b}}{(1+\mathcal{C}_{b})^{2}+\mathcal{C}^{'2}_{b}}\right)(\Delta\omega)^{2}
\eea
Considering only the leading order contribution to the fidelity, we get,
\bea
\label{23}
F_\text{CJ} & = & 1-\frac{(1+\mathcal{C}_{b})}{(1+\mathcal{C}_{b})^{2}+\mathcal{C}^{'2}_{b}}-\frac{N\mathcal{C}\Gamma^{2}_{e}}{|\Omega/2|^{4}}(\Delta\omega)^{2}- \left(\frac{1}{\kappa}+\frac{N\mathcal{C}\Gamma_{e}}{|\Omega/2|^{2}}\right)^{2}(\Delta\omega)^{2}
\\\nonumber
\eea

For the dual-rail case, we calculate a conditional swap fidelity 
\eqref{entanglement_swap_fidelity_jj_final} and the success probability 
\eqref{P_suc_definition}. We can write 
\eqref{entanglement_swap_fidelity_jj_final} as
\bea
\label{25a}
F_\text{swap} = \frac{1}{16P_\text{suc}}\left(4+|\Delta\mathcal{R}|^{2}+4\mathbf{Re}[\Delta\mathcal{R}]+|\Delta\mathcal{R}'|^{2}(\Delta\omega)^{2}
+2\mathbf{Re}[\Delta\mathcal{R}''](\Delta\omega)^{2}+\mathbf{Re}[\Delta\mathcal{R}\Delta\mathcal{R}''^\ast](\Delta\omega)^{2}\right)
\eea
which under the assumption that $\mathcal{C}_{b}, \mathcal{C}'_{b} \gg 1$ and $\mathcal{C}^{\ast\beta}_{b}, \mathcal{C}^{\ast\alpha}_{b} < \mathcal{C}^{2}_{b}$, becomes,
\bea
& = & \frac{1}{16P_\text{suc}}\Bigg(16\left[1 -\frac{(1+\mathcal{C}_{b})}{(1+\mathcal{C}_{b})^{2}+\mathcal{C}^{'2}_{b}}\right]+\frac{4}{(1+\mathcal{C}_{b})^{2}+\mathcal{C}^{'2}_{b}}-\frac{8N\mathcal{C}\Gamma^{2}_{e}}{|\Omega/2|^{4}}(\Delta\omega)^{2}\left(1+\frac{\mathcal{C}_{b}(1+\mathcal{C}_{b})+\mathcal{C}^{'2}_{b}}{(1+\mathcal{C}_{b})^{2}+\mathcal{C}^{'2}_{b}}\right)\nonumber\\
&-&8\left(\frac{1}{\kappa}+\frac{N\mathcal{C}\Gamma_{e}}{|\Omega/2|^{2}}\right)^{2}\left(1+\frac{\mathcal{C}_{b}(1+\mathcal{C}_{b})+\mathcal{C}^{'2}_{b}}{(1+\mathcal{C}_{b})^{2}+\mathcal{C}^{'2}_{b}}\right)(\Delta\omega)^{2}+4\left(\frac{1}{\kappa}+\frac{N\mathcal{C}\Gamma_{e}}{|\Omega/2|^{2}}\right)^{2}(\Delta\omega)^{2}\nonumber\\
&-&8\left(\frac{1}{\kappa}+\frac{N\mathcal{C}\Gamma_{e}}{|\Omega/2|^{2}}\right)\left(\frac{1}{\kappa}+\frac{N^\alpha_{EIT}\mathcal{C}\Gamma_{e}}{|\Omega/2|^{2}}\right)\frac{1}{(1+\mathcal{C}_{b})^{2}}(\Delta\omega)^{2}\Bigg)
\eea
The success probability \eqref{P_suc_definition} is then
\bea
\label{25b}
P_\text{suc} = \frac{1}{4} \Big(2+|\mathcal{R}_{g}|^{2}+|\mathcal{R}'_{g}|^{2}(\Delta\omega)^{2}+\mathbf{Re}[\mathcal{R}_{g}\mathcal{R}''^\ast_{g}](\Delta\omega)^{2}
+|\mathcal{R}_{k}|^{2}+|\mathcal{R}'_{k}|^{2}(\Delta\omega)^{2}+\mathbf{Re}[\mathcal{R}_{k}\mathcal{R}''^\ast_{k}](\Delta\omega)^{2}\Big),
\eea
which under the assumption that $\mathcal{C}_{b}, \mathcal{C}'_{b} \gg 1$ and $\mathcal{C}^{\ast\beta}_{b}, \mathcal{C}^{\ast\alpha}_{b} < \mathcal{C}^{2}_{b}$, becomes,
\bea
\label{26}
P_\text{suc} =
1-\frac{\mathcal{C}_{b}}{(1+\mathcal{C}_{b})^{2}+\mathcal{C}^{'2}_{b}}-\frac{N\mathcal{C}\Gamma^{2}_{e}}{|\Omega/2|^{4}}(\Delta\omega)^{2}
\eea
Substituting Eq. (\ref{26}) and (\ref{22}) into Eq. (\ref{entanglement_swap_fidelity_jj_final}), we get the expression for the conditional swap fidelity, 
\bea
\label{27}
F_\text{swap} &\simeq& 1-\frac{3}{4[\mathcal{C}^{2}_{b}+\mathcal{C}^{'2}_{b}+2\mathcal{C}_{b}]}-\frac{\mathcal{C}^{2}_{b}}{[\mathcal{C}^{2}_{b}+\mathcal{C}^{'2}_{b}+2\mathcal{C}_{b}]^{2}}-\frac{3}{4}\left[\frac{1}{\kappa}+\frac{N\mathcal{C}\Gamma_{e}}{|\Omega/2|^{2}}\right]^{2}\left[1+\frac{\mathcal{C}_{b}}{\mathcal{C}^{2}_{b}+\mathcal{C}^{'2}_{b}+2\mathcal{C}_{b}}\right](\Delta\omega)^{2}\nonumber\\
\eea
Finally, keeping only the dominant contribution to the gate operation, we get the conditional swap fidelity,
\bea
\label{28}
F_\text{swap} = 1-\frac{1}{[\mathcal{C}^{2}_{b}+\mathcal{C}^{'2}_{b}]}-\frac{3\mathcal{C}^{2}_{b}-\mathcal{C}^{'2}_{b}}{4[\mathcal{C}^{2}_{b}+\mathcal{C}^{'2}_{b}]^{2}}-\frac{3}{4}\left[\frac{1}{\kappa}+\frac{N\mathcal{C}\Gamma_{e}}{|\Omega/2|^{2}}\right]^{2}(\Delta\omega)^{2}
\eea

\section{Inhomogeneous Ensemble}
So far, we have considered only a homogeneous ensemble without decay of the Rydberg level. In this section we discuss the case for an inhomogeneous ensemble. For simplicity, we only consider $\Delta\omega = 0$. Here the scattering dynamics depends on where the excitation was stored in the ensemble. From the fidelity expressions Eq.~\eqref{CJ_fidelity_single_rail} and Eq.~\eqref{entanglement_swap_fidelity_jj_final} we see that the essential parameter is $\sum_{k}|\alpha_{k}|^{2} \mathcal{R}_{k}$ which for $\delta_{l} = \Delta_{l} = 0$ is given by
\bea
\label{29}
\sum_{k}|\alpha_{k}|^{2}\mathcal{R}_{k} =  \sum_{k}|\alpha_{k}|^{2}\left[\frac{2}{1+\sum_{l}\frac{|\mathcal{G}_{l}|^{2}/\kappa\Gamma_{el}}{1+|\Omega_{l}/2|^{2}/(\Gamma_{rl}\Gamma_{el}+i\mathcal{V}_{kl}\Gamma_{el})}}-1\right].
\eea
When the ensemble was homogeneous, we defined the blockaded co-operativity $\mathcal{C}_{b}$ and $\mathcal{C}'_{b}$ such that $\mathcal{R}_{k} = \frac{1}{1+\mathcal{C}_{b}+i\mathcal{C}_{b}}$.  Analogous to this for an inhomogeneous ensemble, we can define the blockaded co-operativity through $\sum_{k}|\alpha_{k}|^{2}\mathcal{R}_{k} = 1/(1+\mathcal{C}^\text{inh}_{b}+i\mathcal{C}^{' \text{inh}}_{b})$ where, 
\bea
\label{20a}
\mathcal{C}^\text{inh}_{b} & = &\mathbf{Re}\left[\frac{1}{\sum_{k}|\alpha_{k}|^{2}\left(1+\sum_{l\neq k}\frac{\mathcal{C}_{l}}{1+|\Omega/2|^{2}/(\Gamma_{rl}\Gamma_{el}+i\mathcal{V}_{kl}\Gamma_{el})}\right)^{-1}}\right]-1,\nonumber\\
\mathcal{C}^{' \text{inh}}_{b} & = & \mathbf{Im}\left[\frac{1}{\sum_{k}|\alpha_{k}|^{2}\left(1+\sum_{l\neq k}\frac{\mathcal{C}_{l}}{1+|\Omega/2|^{2}/(\Gamma_{rl}\Gamma_{el}+i\mathcal{V}_{kl}\Gamma_{el})}\right)^{-1}}\right] 
\eea
Note that contrary to the homogeneous case the above defined effective co-operativity for inhomogeneous ensemble also includes the effect of Rydberg decoherence on the scattering process. Thus, to study the Fidelity of the phase gate for an inhomogeneous ensemble and in presence of decoherence, the results in Eqs. (\ref{23}) and (\ref{28}) can be utilized but now with $\mathcal{C}_{b}$ replaced by $\mathcal{C}^\text{inh}_{b}$ and $\mathcal{C}'_{b}$ by $\mathcal{C}^{' \text{inh}}_{b}$.

\end{document}